\documentclass[aps,prb,twocolumn,showpacs,superscriptaddress,groupedaddress,floatfix]{revtex4-1}
\usepackage{latexsym}
\usepackage{graphicx}
\usepackage{dcolumn}
\usepackage{bm}
\usepackage{amssymb}
\usepackage{amsmath}
\usepackage[space]{grffile}
\usepackage{subcaption}
\usepackage[usenames,dvipsnames]{color}

\newcommand{\gs}{|0\rangle}

\newcommand{\Tr}{\text{Tr}}

\newcommand{\bra}[1]{\langle #1 |}
\newcommand{\ket}[1]{|#1\rangle}

\begin{document}

\title{Entanglement properties of the time periodic Kitaev Chain}

\date{\today}

\begin{abstract}
The entanglement properties of the time periodic Kitaev chain with nearest neighbor and next nearest neighbor hopping, is studied. The cases of
the exact eigenstate of the time periodic Hamiltonian, referred to as the Floquet ground state (FGS),
as well as a physical state obtained from time-evolving an initial state unitarily under the influence of the time
periodic drive are explored. Topological phases are characterized by different numbers of Majorana zero ($\mathbb{Z}_0$)
and $\pi$ ($\mathbb{Z}_{\pi}$) modes, where the zero modes are present even in the absence of the drive, while the $\pi$ modes
arise due to resonant driving.
The entanglement spectrum (ES) of the FGS as well as the physical state show topological Majorana modes whose number is different from
that of the quasi-energy spectrum. The number of Majorana edge modes in the ES of the FGS vary in time from
$|\mathbb{Z}_0-\mathbb{Z}_{\pi}|$ to $\mathbb{Z}_0+\mathbb{Z}_{\pi}$ within one drive cycle, with the maximal $\mathbb{Z}_0+\mathbb{Z}_{\pi}$ modes appearing
at a special time-reversal symmetric point of the cycle.
For the physical state on the other hand, only the modes inherited from the initial
wavefunction, namely the $\mathbb{Z}_0$ modes, appear in the ES. The $\mathbb{Z}_{\pi}$
modes are absent in the physical state as they
merge with the bulk excitations that are simultaneously created due to resonant driving.
The topological properties of the Majorana zero and $\pi$ modes in the ES are also explained by mapping the parent wavefunction to a Bloch sphere.
\end{abstract}

\author{Daniel J. Yates}

\author{Aditi Mitra}

\affiliation{Department of Physics, New York University, 726 Broadway, New York, NY, 10003, USA}
\maketitle
\section{Introduction}

Topological insulators (TIs) and metals are now a major part of condensed matter research~\cite{Hasan10,Zhang11,Felser17}.
While traditional topological systems such as integer~\cite{Klitzing80} and fractional~\cite{Stormer82} quantum Hall states have a clear
transport signature, the consequences
of topology for almost all other topological systems is far more subtle. The bulk-boundary correspondence implies
protected edge states, however these states do not have as dramatic a signature on transport as quantum Hall systems.
This has lead to more creative ways to identify topological systems in the laboratory, such as via direct probe of
edge states by ARPES~\cite{Hasan17}, or more sophisticated proposals for exhibiting braiding statistics from
exchanging edge modes~\cite{Kitaev06}.

A new class of TIs are those that arise due to periodic driving~\cite{Oka09,Kitagawa10,Kitagawa11,Lindner11}.
Almost all topological insulators and metals are argued to have a time-periodic version~\cite{Sondhi16,Sondhi16b,Else16,Vishwanath16,Po16,Roy16,Roy17b,Roy17}.
The manifestation of topology in these systems is even more complex~\cite{Kundu13,Dehghani14,Dehghani15a,Dehghani16}
because for an out of equilibrium system, the state may be far from an exact eigenstate of the Hamiltonian,
and coupling to a low temperature reservoir does not always ensure the appearance of a Gibbs' state~\cite{Kohn09,Dehghani15b,Lindner15,Dehghani16b}.

Due to all this, an appealing way of characterizing the topology, that does not rely on specific transport signatures, nor assumptions
requiring the system to be in thermal equilibrium, is via
a study of the wavefunction or the system density matrix itself. This can be done by employing various information theoretic measures.
For wavefunctions that are ground states of static Hamiltonians, the entanglement entropy (EE) shows
area law due to the gapped spectrum, much like ground state wavefunctions of generic gapped Hamiltonians. This is not
so useful from the point of view of topology, barring a few exceptions where the EE shows subleading corrections due to
topology~\cite{Levin06,Preskill06,Hamma13}.
In contrast, the bulk-boundary correspondence for
the spectrum of the Hamiltonian remarkably also translates to a bulk boundary correspondence in the spectrum of the
reduced density matrix, which is referred to as the entanglement spectrum (ES)~\cite{Haldane08,Fidkowski10}.
A natural question to ask is, how does the bulk boundary correspondence manifest in the entanglement properties of time-periodic systems?
This issue has been addressed for Floquet Chern insulators~\cite{Yates16}.
In this paper, we address this for the time-periodic Kitaev chain~\cite{Kitaev01}.

We study a Kitaev chain representing a mean-field $p$-wave superconductor, and allow for nearest-neighbor (NN)
and next nearest-neighbor (NNN) hopping. The chemical potential is made to vary periodically in time.
In the absence of a periodic drive, the static Kitaev chain preserves time-reversal and particle hole
symmetry, and falls into the BDI class of the Altland Zirnbauer (AZ)  classification~\cite{Altland97,Ryu10}.
The topological invariant is an integer $\mathbb{Z}$ corresponding to
$\mathbb{Z}$ Majorana zero modes. Periodic drive makes the ``energy'' or quasi-energy spectrum periodic, and
elevates the topological invariant to~\cite{Vishwanath16} $\mathbb{Z}\times \mathbb{Z}$,
where the first $\mathbb{Z}$
represent the number of zero quasi-energy Majorana modes (MZM), while the second integer represents $\mathbb{Z}$ Majorana modes
at the Floquet zone boundaries, the so called Majorana $\pi$ modes (MPM).

In this paper we explore the fate of the MZM and MPM on the ES. We study two states, one
is the exact eigenstate of the time-periodic Hamiltonian, also known as the Floquet mode.
We refer to this state as the Floquet Ground State (FGS). The second state which we study is one that is
the ground state of the static Kitaev chain, but unitarily time-evolved under the influence of a time-periodic
drive. We refer to this as the physical or quenched state as the dynamics is under the influence
of a rapid switch on protocol of the periodic drive.

The new topological features of the drive arise entirely due to resonant band-crossings, resulting in MPMs.
We define a resonant process as one where the frequency of the drive is such that an on-shell processes can connect the ground state to the excited state
of the static Hamiltonian. Since there is no adiabatic limit for a resonant drive~\cite{Privitera16,Yates16},  the main physics we uncover, namely which
topological modes are present in the ES of the physical state and which are absent, does not depend on how fast the drive has been switched on.

The paper is organized as follows. Section~\ref{secM} introduces the model and outlines construction of the
Floquet Ground State (FGS), and the physical state arising from a quench.  This section also presents the quasi-energy spectrum and
quasi-modes for a finite wire with open boundary conditions, in order to highlight the appearance of MZM and MPM modes
in the physical boundary. This will form a helpful point of comparison with the ES and the Schmidt states that reside at the entanglement cut.

Section~\ref{secE} discusses the construction of the entanglement Hamiltonian.
Section~\ref{secG} presents results for the entanglement properties for the ground state wavefunction of the static Hamiltonian,
including an analytic solution for the MZM on the entanglement cut.
Section~\ref{secF} presents the entanglement properties for the FGS. The main features of the ES of the FGS are elucidated using a
spinor representation in Section~\ref{secS}.
Finally section~\ref{secQ} presents the entanglement properties for the physical state obtained from unitary time evolution under the
effect of the periodic drive. We present our conclusions in Section~\ref{secD}.

\section{Model} \label{secM}

Our Hamiltonian is the Kitaev model, with the addition of NNN hopping,
\begin{align}
	\begin{split} \label{ham}
	H &= \sum_i -t_h \left( c_i^\dagger c_{i+1} + c_{i+1}^\dagger c_i \right) \\
	&\qquad  -\Delta \left( c_i^\dagger c_{i+1}^\dagger +c_{i+1} c_{i} \right)
	-\mu(t) \left( c_{i}^\dagger c_i - \frac{1}{2}\right) \\
	&\qquad -t_h'\left(c_i^\dagger c_{i+2} + c_{i+2}^\dagger c_{i}\right)
	-\Delta'\left( c_i^\dagger c_{i+2}^\dagger + c_{i+2} c_{i} \right).
	\end{split}
\end{align}
Unless otherwise stated, results will be presented with NNN hopping turned off ($\Delta' = t_h' = 0$) except in Section~\ref{NNNsec}.
We drive the system with a periodic chemical potential,
\begin{equation*}
\mu(t) = \mu_0 + \xi \sin (\Omega t).
\end{equation*}
The system is symmetric under $\mu \rightarrow - \mu$; we enforce $\mu_0 \ge 0$.
Working with the static case for now, $H(t) \rightarrow H(\xi=0)$, we can diagonalize the system with a
Fourier transform,
\begin{equation}\label{ham2}
	H(\xi=0) =\sum_{k}
	\begin{pmatrix}
	c_k^\dagger & c_{-k}
	\end{pmatrix}
	H_{\rm BdG}(k)
	\begin{pmatrix}
	c_k \\ c_{-k}^\dagger
	\end{pmatrix} ,
\end{equation}
where $H_{\rm BdG}(k)$ is the Bogoliubov-de-Gennes (BdG) Hamiltonian,
\begin{align*}
H_{\rm BdG}(k) &=
-\left(\Delta \sin (k)+ \Delta' \sin(2k)\right) \sigma_y \\
&\qquad - \left( t_h \cos (k) + t_h' \cos(2k) + \frac{\mu}{2} \right) \sigma_z\\
&= \vec{h}_k \cdot \vec{\sigma},
\end{align*}
where the momenta $k$ are in units of the lattice spacing.

\subsection{Ground State Wave function of static Hamiltonian}

The static BdG Hamiltonian can be fully diagonalized via a Bogoliubov transformation,
\begin{equation}
	\begin{pmatrix}
		d_k \\ d_{-k}^\dagger
	\end{pmatrix}
	=
	\begin{pmatrix}
		u & v \\ -v^* & u^*
	\end{pmatrix}
	\begin{pmatrix}
		c_k \\ c_{-k}^\dagger
	\end{pmatrix},
\end{equation}
where we let $u = \cos \theta_k $, $v = -i \sin \theta_k$, and
\begin{equation}\label{eq:theta}
\theta_k = \frac{1}{2} \arctan \left(\frac{\Delta \sin(k)}{t_h\cos (k) + \frac{\mu}{2}}\right).
\end{equation}
The result of the transformation is,
\begin{align*}
	\label{Hamiltonian3}
	H(\xi=0) &= \sum_{k}
\text{sgn}\left[h_{k,z}\right]\epsilon_k \left(d_k^\dagger d_k - d_{-k}d_{-k}^\dagger \right)\\
&= \sum_k \text{sgn}\left[h_{k,z}\right] 2 \epsilon_k d_k^\dagger d_k,
\end{align*}
and $\epsilon_k$ is
\begin{equation*}
\epsilon_k = \sqrt{\left( t_h \cos k + \frac{\mu_0}{2} \right)^2 + \Delta^2 \sin^2 k}.
\end{equation*}
We are free to redefine $d_k^\dagger \leftrightarrow d_k$ for each $|k|$ sector to 
ensure that $\text{sgn}\left[ h_{k,z}\right] \epsilon_k>0$, resulting in,
\begin{equation}
	H(\xi=0) = \sum_k 2\epsilon_k d_k^\dagger d_k.
\end{equation}
The ground state is the dressed vacuum,
\begin{align*}
\ket{\text{GS}} &= \prod_{k} d_k \ket{0} \\
&= \prod_{k>0} \left[d_{-k}d_k\right] \ket{0}.
\end{align*}
We can also express this in terms of the original operators.
When $h_{k,z}>0$, the ground state is:
\begin{equation} \label{eq:gs1}
	|\text{GS} \rangle_{|k|} = \left[i \sin \theta_k c_k^\dagger c_{-k}^\dagger\gs + \cos \theta_k \gs \right],
\end{equation}
and when $h_{k,z}<0$:
\begin{equation} \label{eq:gs2}
	|\text{GS} \rangle_{|k|} = \left[\cos \theta_k c_k^\dagger c_{-k}^\dagger \gs+ i \sin \theta_k \gs \right].
\end{equation}
Considering these states as vectors in the $\left(c_k^\dagger c_{-k}^\dagger\gs, \gs\right)$ basis,
the ground state at each point $|k|$ is simply the numerical ground state of the BdG Hamiltonian.

\subsubsection{Topology of the static system}

We discuss the anti-unitary symmetries present in our model.
The particle-hole symmetry (PHS) is manifest in \eqref{ham2},
$\sigma_x H_{\rm BdG}^*(-k) \sigma_x = -H_{\rm BdG}(k)$.
Physically, this corresponds to the excitation energy of a quasi-particle being equivalent to the
addition of a quasi-hole.  Our model also
has time reversal symmetry (TRS). This corresponds to complex conjugation $\mathcal{K}$ for spinless fermions,
and is clear in the position space definition of the model \eqref{ham}.
While TRS may be a confusing issue for spin-less fermions, note that the model can also be derived
from a spin-chain where the spinless fermions of our model are simply the Jordan-Wigner fermions~\cite{Niu12, Sen13}.
With both PHS and TRS, our model falls into the BDI Altland Zirnbauer (AZ) class~\cite{Altland97, Ryu10}, which for one dimension,
has a topological index of $\mathbb{Z}$. This index counts the number of Majorana edge modes.

Writing $H_{\rm BdG}=\vec{h}_k\cdot\vec{\sigma}$, PHS imposes $h_x(k)=-h_x(-k), h_y(k)=-h_y(-k), h_z(k)=h_z(-k)$.
In addition, TRS corresponds to invariance under complex conjugation and $k\rightarrow -k$, implying $h_x=0$.
This is equivalent to the statement that the
vector $\vec{h}_k$ has no $\sigma_x$ component and thus lies in the $\sigma_{y,z}$ plane. It therefore
admits a well defined winding number definition via the pseudo-vector $\vec{h}_k$ encircling the origin.
The winding number can  be conveniently extracted
by writing the eigenstates of our two level system
at each $k$ as a spinor $(\cos\alpha/2, e^{i \beta} \sin \alpha/2)$, which rests on the
unit sphere parameterized by $\alpha$ and $\beta$.
The topological state consists of the spinors
in the Brillouin zone connecting the north and south poles. We will use this spinor analogy later when we discuss
the topological features in the ES of the FGS.

When one considers only a single wire with no NNN terms ($t_h'=0,\Delta'=0$), the original construction of the
Kitaev model\cite{Kitaev01} only considered a $\mathbb{Z}_2$ invariant consistent with class D~\cite{Altland97,Ryu10}
(PHS but no TRS). The fact that it is now considered to
be in class BDI  appears contradictory, but turns out to not matter as the topological indices are $\pm 1,0$ when only NN hopping is present.
So for the purposes of counting edge states,
$\mathbb{Z}$ and $\mathbb{Z}_2$ will predict the same number and there is no contradiction.

In our paper we specifically would like to lift the ambiguity between BDI and D. Therefore we also present results with NNN
hopping in section~\ref{NNNsec}. This longer range hopping generates more Majorana edge states clearly placing our model in the BDI category. Consequently we will discuss
the stability of the observed edge modes in the ES  to TRS and PHS preserving perturbations.

\subsection{Floquet Ground State}

We now consider the time-dependent problem, and hence restore the time-dependence of the chemical potential in Eq.~\eqref{ham}.
For the time-periodic system we solve the Schr\"{o}dinger equation with the Floquet ansatz $\ket{\psi(t)} = e^{-i \epsilon_a t}\ket{a(t)}$,
where $\ket{a(t+T)} = \ket{a(t)}$ is time periodic, and $\epsilon_a$ are the quasi-energies.
We call the states $\ket{\psi(t)}$ the Floquet modes, they are eigenstates of the propagator when time evolved
an integer number of periods. We find the quasi-energies and quasi-modes by solving the Floquet-Bloch equation
\begin{equation*}
\left[ H_k(t) - i\partial_t \right] \ket{a_k(t)} = \epsilon_a \ket{a_k(t)}.
\end{equation*}
The operator $H_F=H_k(t) - i\partial_t$ is termed the Floquet Hamiltonian.

\begin{figure}
\includegraphics[width = .95\linewidth, keepaspectratio]{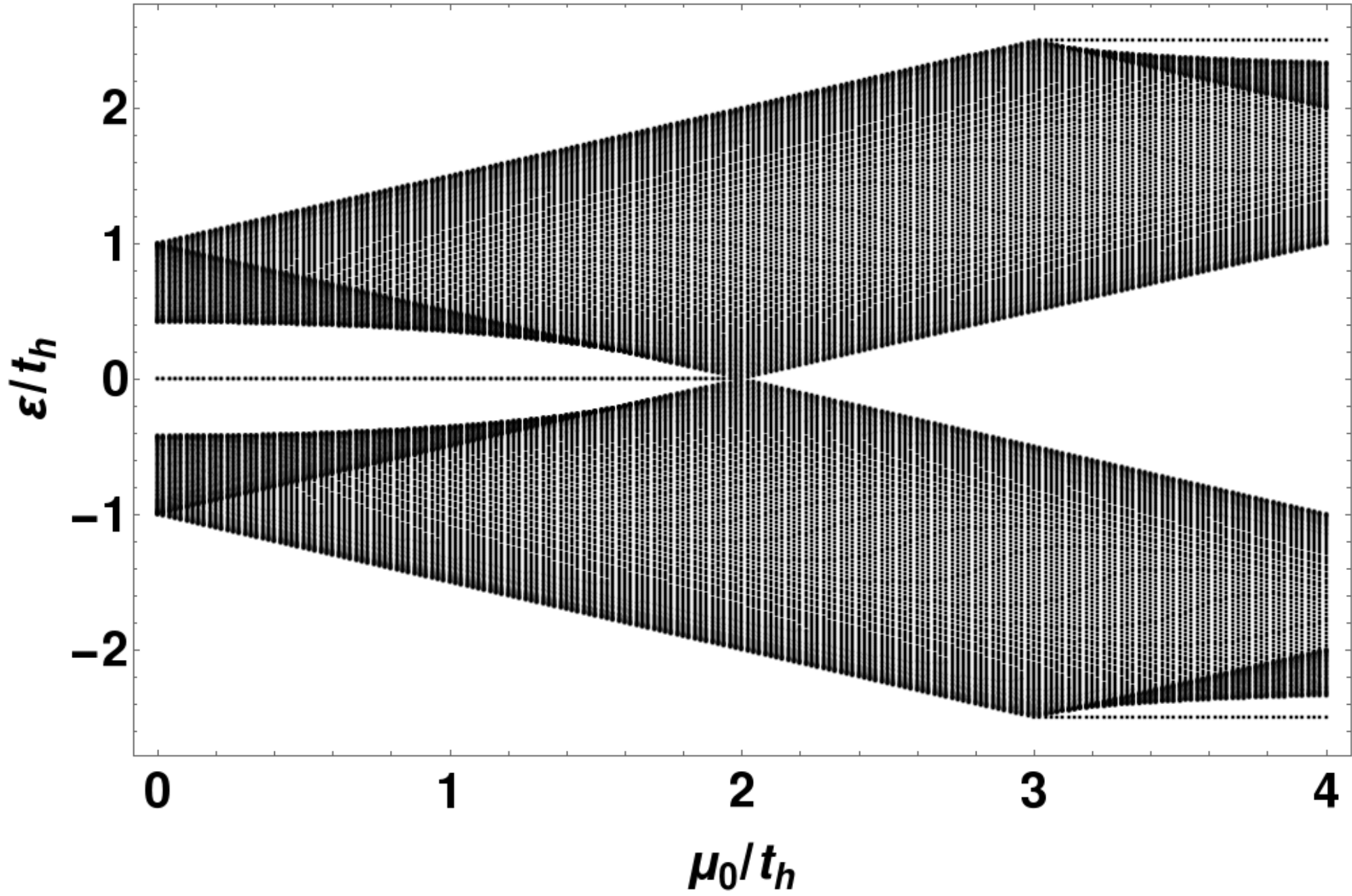}
\caption{Quasi-energy levels as a function of $\mu_0/t_h$ for a driven, finite wire of 100 sites.
Here $\Delta/t_h = .5$, $\xi/t_h = 2$, $\Omega/t_h = 5$.
The phases where $\mu_0/t_h<3$ correspond nearly exactly with the energy levels of a static finite wire.
When $\mu_0/t_h\approx 3$, the resonance condition drastically
modifies the structure of the quasi-energy bands around the FBZ boundaries, with the appearance of
Majorana $\pi$ modes (MPMs) at $\epsilon/t_h \approx \pm \Omega/2$. These modes are separated from the bulk band.
From left to right, we have a MZM phase ($0<\mu_0/t_h<2$), trivial phase ($2<\mu_0/t_h<3$), and a MPM phase ($3<\mu_0/t_h<4$).
}
\label{finite_om5}
\end{figure}

\begin{figure}
\includegraphics[width = .95\linewidth, keepaspectratio]{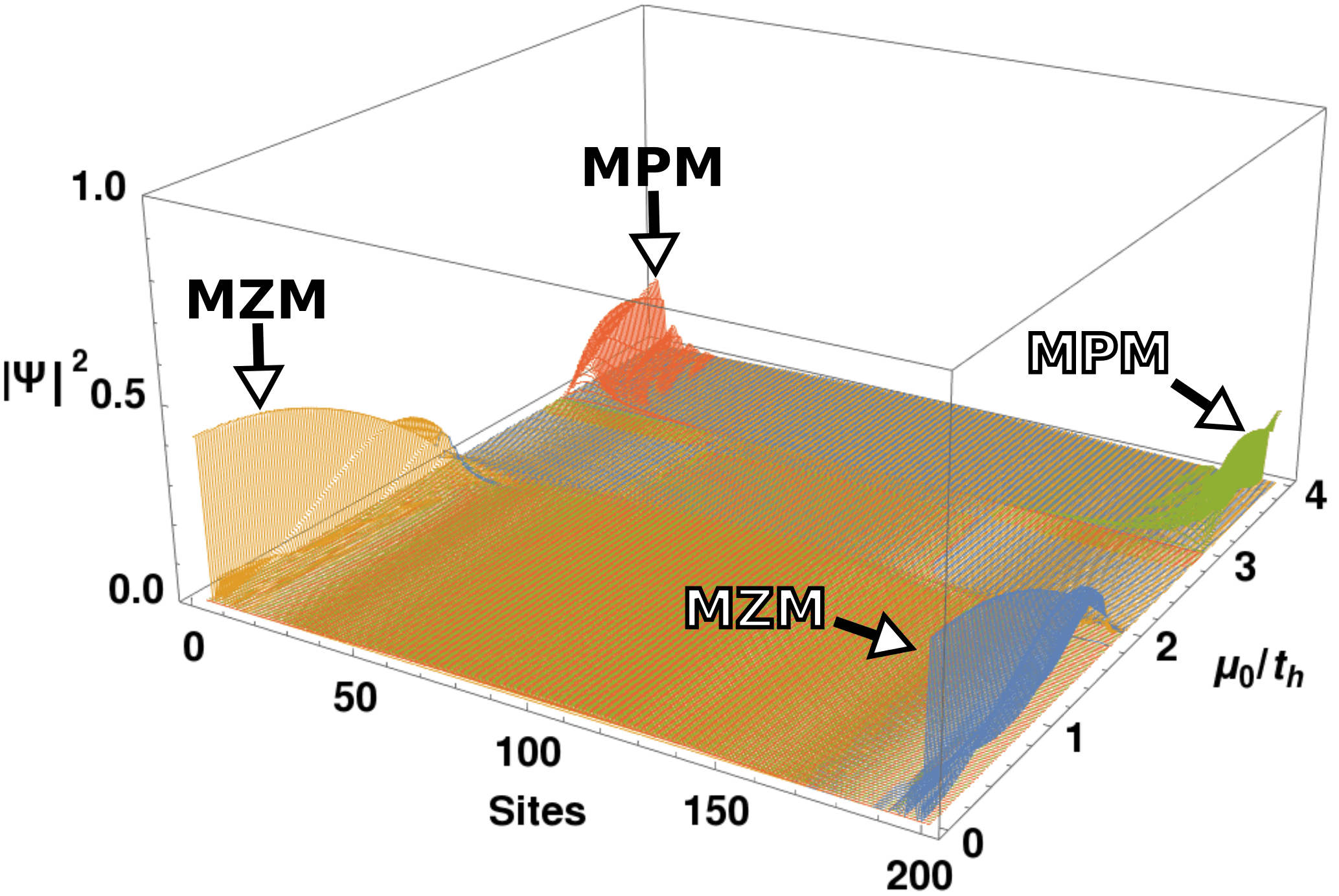}
\caption{
Edge states verses $\mu_0/t_h$ of a driven, finite wire of a 100 sites.
Here $\Delta/t_h = .5$, $\xi/t_h = 2$, $\Omega/t_h = 5$.
The MZM and MPM states are localized on the left and right ends of the wire. Distinct colors (red, orange, green, blue) denote distinct eigenstates.
We note that the Schmidt states
have double the number of sites because these states are in the Majorana basis. This
doubling is equivalent to the effective spinor degree of freedom describing the superconductor.
}
\label{finite_om5_edges}
\end{figure}

\begin{figure}
\includegraphics[width=.95\linewidth]{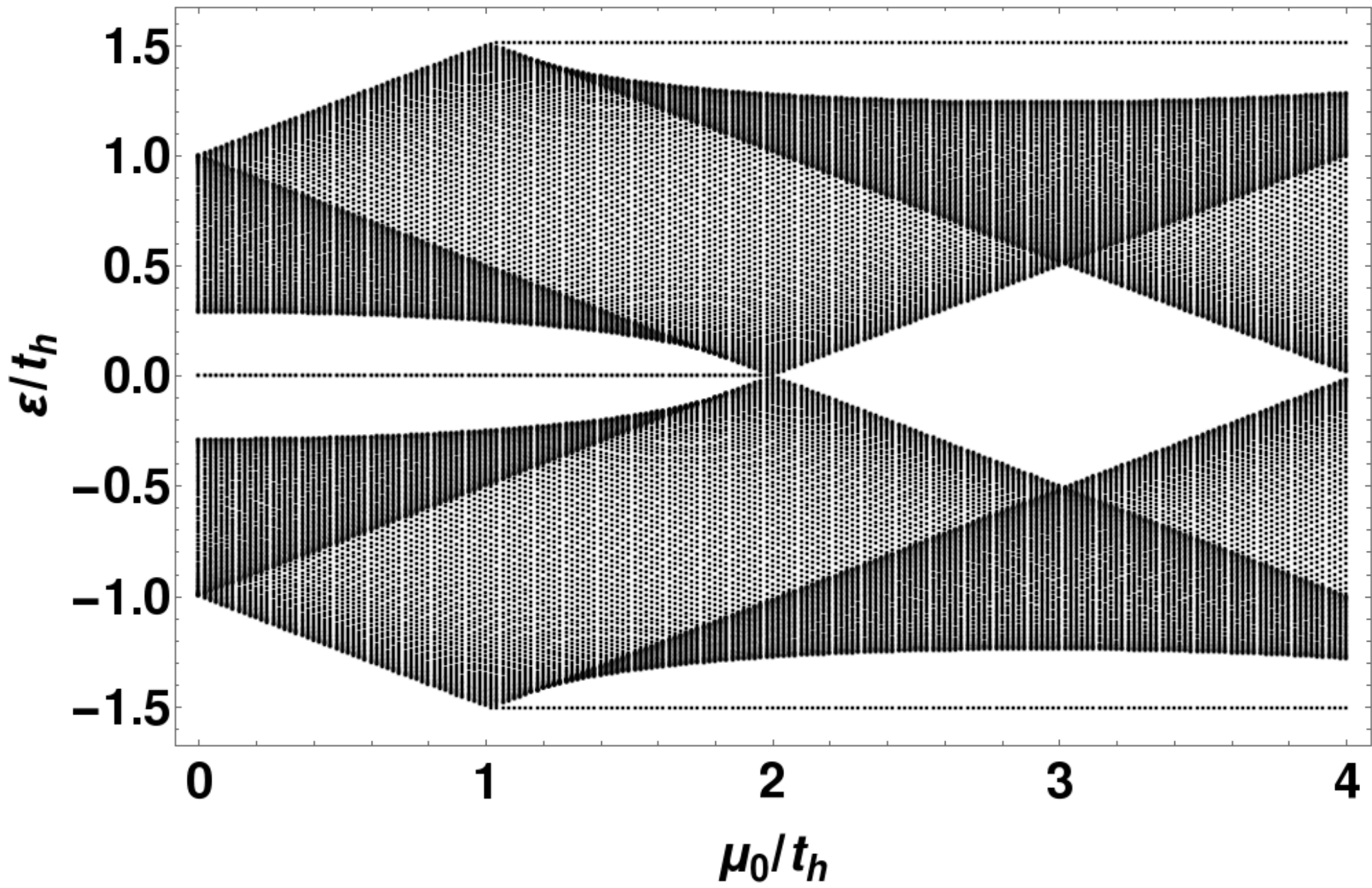}
\caption{Quasi-energy levels as a function of $\mu_0/t_h$ for a driven, finite wire of 100 sites.
Here $\Delta/t_h = .5$, $\xi/t_h = 2$,
$\Omega/t_h = 3$.
The resonance condition occurs for smaller values of $\mu_0/t_h \approx 1$ relative to the example in figure~\ref{finite_om5}.
The resonance occurs for values of $\mu_0$ where the static wire is topological.
Thus from left to right, we have a MZM phase ($0<\mu_0/t_h<1$), a MZM\&MPM phase ($1<\mu_0/t_h<2$), and a MPM phase ($2<\mu_0/t_h<4$).
}
\label{finite_om3}
\end{figure}

\begin{figure}
\includegraphics[width = .95\linewidth, keepaspectratio]{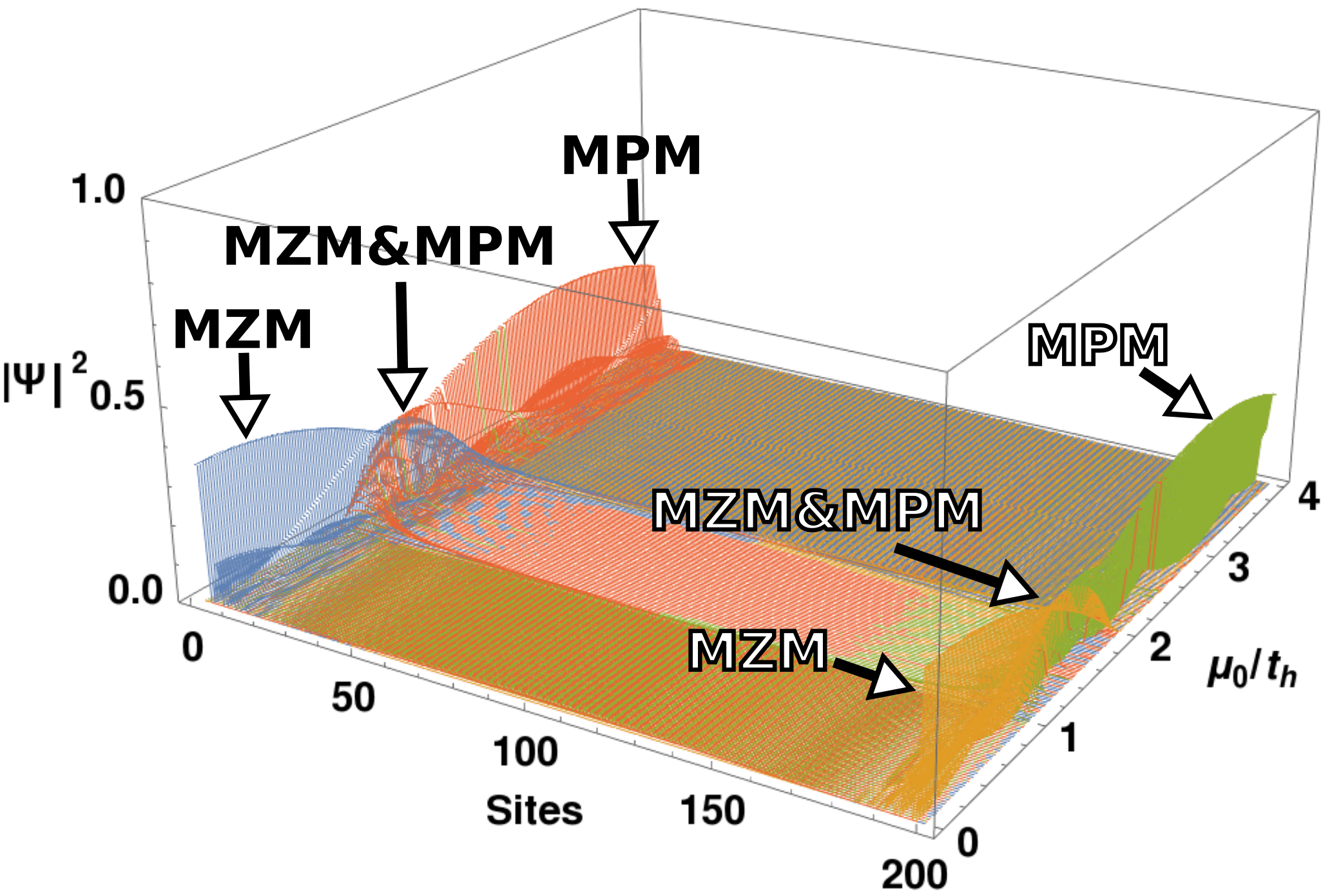}
\caption{
Edge states verses $\mu_0/t_h$ of a driven, finite wire of a 100 sites. Here $\Delta/t_h = .5$, $\xi/t_h = 2$, $\Omega/t_h = 3$.
The MZM and MPM states live on the left or right ends of the wire. Distinct colors (red, orange, green, blue) denote distinct eigenstates.
In the MZM\&MPM phase both a MZM and MPM live at each end of the wire.
The MZM\&MPM states are localized to within a few sites of the edge of the wire. Even though MZM\&MPM modes spatially overlap, they
do not couple as they do not share the same quasi-energy. They will however couple in the ES.
}
\label{finite_om3_edges}
\end{figure}

In order to calculate the Floquet modes, we expand the Floquet Hamiltonian into the
expanded Hilbert space\cite{Sambe73,Eckardt15} $\mathcal{H} \otimes \mathcal{L_T}$, where 
$\mathcal{L}_T$ is the space of periodic functions.
The result of the expansion is,
\begin{equation*}
H_F(k) \equiv
\sum_{n,m} \left[ \frac{1}{T} \int_0^T dt' H_{k}(t') e^{-i (n-m)\Omega t'} + \delta_{n,m} m\Omega \right].
\end{equation*}
This leads to a time independent matrix indexed by the photon numbers $n,m$.
The elements of the matrix are various frequency expansions of the original time-dependent Hamiltonian.
Denoting the $m$-th Fourier expansion of
$H_k(t)$ as $H_k^{(m)}=\frac{1}{T} \int_0^T dt' H_{k}(t') e^{-i m \Omega t'}$,
for our model, the matrix is triple-banded, with the center-most bands for a given row $n$, from left to right being,
$H_k^{(-1)}$, $H_k^{(0)} + n \Omega$, and $H_k^{(1)}$, where
$H_k^{(\pm 1)}= \frac{\pm i \xi \sigma_z}{4}$ and $H_k^{(0)}$ is the static Hamiltonian.

In the limit of a weak drive ($\xi/t_h\ll 1$), and a highly off-resonant frequency  much larger
than the bandwidth of $H_0$ ($\Omega/t_h\gg 1$), the spectrum of the expanded Floquet Hamiltonian will largely be copies of the
static Hamiltonian repeated at integer multiples of $\Omega$.
The expanded Floquet Hamiltonian contains a large amount of redundancies,
as the majority of the eigenstates from its spectrum will produce the same Floquet modes in the traditional
Hilbert space. To avoid over-counting, we restrict ourselves to the Floquet Brillouin Zone (FBZ) of the eigenstates contained
within the quasi-energy range $\epsilon_a<|\Omega/2|$.

The diagonalization in the extended space
effectively creates a time-periodic unitary (Bogoliubov) transformation
on our initially static Nambu spinor.  This transformation will diagonalize our
Floquet Hamiltonian and result in the following, where the time dependence is
absorbed into the new operators,
\begin{equation*}
2\epsilon_k \left(\tilde{d}_k^\dagger(t)\tilde{d}_k(t) + \tilde{d}_{-k}^\dagger(t) \tilde{d}_{-k}(t) -1\right) \ket{a_k(t)}
= \epsilon_k \ket{a_k(t)}.
\end{equation*}

There are four possible eigenstates $\ket{\text{FGS}} = \tilde{d}_{-k}\tilde{d}_k \gs$,
$\tilde{d}_k^\dagger \ket{\text{FGS}}$, $\tilde{d}_{-k}^\dagger\ket{\text{FGS}}$, and $\tilde{d}_{-k}^\dagger
\tilde{d}_k^\dagger\ket{\text{FGS}}$.
We can disregard the odd-parity states here as our ground state is even-parity i.e., we allow
for only doubly occupied or empty sites. These states are equivalent to finding the two component (numerical) eigenvectors in the basis,
$\left( c_k^\dagger c_{-k}^\dagger |0\rangle, |0\rangle \right)$, we denote the even parity excited state
as $\ket{\text{FES}} = \tilde{d}_k^\dagger \tilde{d}_{-k}^\dagger \ket{\text{FGS}}$.

\subsubsection{Finite wire}

To better understand the bulk-boundary correspondence in the ES, we include here the quasi-energy spectrum and edge states for
physical edges on a finite wire. The spectrum and eigenstates of a driven finite Majorana chain are shown in
figures \ref{finite_om5}, \ref{finite_om5_edges}, \ref{finite_om3}, \ref{finite_om3_edges}.
The calculation of these figures consisted of the tight-binding finite system analogs of the bulk Floquet
quantities outlined above \cite{Sen13,Benito14,Wang14}.

Figures \ref{finite_om5} and \ref{finite_om3} clearly show the appearance of additional $\pi$ edge states that can only
occur in a spectrum that is periodic. Comparing the spectrum with the eigenstates plotted in figures
\ref{finite_om5_edges} and \ref{finite_om3_edges} shows that these
states are indeed edge states. Furthermore, the MPM states in figure \ref{finite_om5} occur in the topologically
trivial region of the static wire, showing that periodic driving can induce topological phase transitions. These topological
phase transitions arise due to resonant gap-closing and re-opening at $\epsilon = \pm\Omega/2$,
with such resonances introducing MPMs into the system.
Figure~\ref{finite_om3} shows that with a drive, one can now have three different scenarios. One is a phase where only MZM exist and
this corresponds to a high frequency off-resonant drive. Second, a
phase where only MPM exist, this corresponds to a resonant drive. And finally a phase where MZM and MPM coexist, also arising
due to a resonant drive.

\subsection{Physical State}
Knowing the Floquet modes, we can construct the propagator:
\begin{equation*}
U_k(t) = \prod_{|k|}\sum_a e^{-i \epsilon_{a,k} t} \ket{a_{k}(t)} \bra{a_k(0)}.
\end{equation*}
The physical time-evolved state we will consider is one which is the ground state of the static Hamiltonian $H(\xi=0)$, but
unitarily time-evolved under the influence of a sudden switch on of the periodic drive at $t=0$. Thus the physical state is,
\begin{align*}
\ket{\Psi(t)} &= \prod_{|k|} U_k(t) \ket{\text{GS}}_k\\
&= \prod_{|k|}\left[
e^{-2i\epsilon_{|k|}t}
\rho_{|k|,\uparrow}\tilde{d}_{-k}^\dagger(t)\tilde{d}_{k}^\dagger(t)\ket{\text{FGS}(t)}_{|k|} \right.\\
&\left. \qquad + e^{2i\epsilon_{|k|}t} \rho_{|k|,\downarrow} \ket{\text{FGS}(t)}_{|k|} \right].
\end{align*}
Where $\rho_{\updownarrow}$ are the time-independent overlaps of the Floquet states with the ground state at the instant the
periodic drive was switched on:
\begin{align}
\rho_{\downarrow,|k|} &= \bra{\text{FGS}_{|k|}(0)}\text{GS}_{|k|}\rangle,\nonumber\\
\rho_{\uparrow,|k|} &= \bra{\text{FES}_{|k|}(0)}\text{GS}_{|k|}\rangle.\label{rhodef}
\end{align}

\subsubsection{Topology of the time-periodic system}

There are two states of interest in the driven setting, the FGS and the quenched state. As
the FGS is the ground state of the effectively time-independent Floquet Hamiltonian, the topology of this state can
be understood via methods analogous to analyzing topologies in conventional static Hamiltonians.
The topology of the quenched state is not as clear to discern and will be discussed later.

For now, using approaches valid for the static case, let us understand the topologies of the Floquet Hamiltonian. For this, we must
identify the anti-unitary symmetries. In the driven setting, it is clear that PHS holds at all times during the drive.
TRS is a more delicate question to answer. One can define TRS for a Floquet system as,
$\mathcal{T}H(\tau - t)\mathcal{T}^{-1} = H(t)$, for some $\tau$ and for all $t$ \cite{Lindner11}.
Such a condition does hold for our drive with $\tau=\pi/\Omega=T/2$. In addition there are two special times during
the drive when the two TRS points coincide {\sl i.e.}, $t=\tau-t, t=\tau+T-t$. For our drive this happens at
$t^*=T/4,3T/4$. We will show later in the paper that the number of Majorana edge modes in the ES show special behavior at $t^*$.

While TRS and PHS hold for the Floquet Hamiltonian, the new feature of Floquet systems is that
we have two gaps, one at $\epsilon = 0$ and the other at the zone boundaries $\epsilon = \pm \Omega/2$. We thus have the possibility of edge
states spanning either or both gaps when the system is placed in a finite geometry.
We expect the BDI classification to persist, so the topological classification in the driven setting becomes~\cite{Vishwanath16}
$\mathbb{Z} \times \mathbb{Z}$. The first integer counts the number of
Majorana edge modes at $\epsilon = 0$ and the second integer counts the number of Majorana
edge modes at $\epsilon = \Omega/2$. Examples showcasing the $\mathbb{Z}\times\mathbb{Z}$
index for the eigenstates are shown in figures \ref{finite_om5} and \ref{finite_om3}.

For the physical or quenched state, the topological features and bulk-boundary correspondence is not
clear. Introducing a physical boundary can non-trivially modify the time-evolution, and create
system dependent excitations. To avoid doing this, we will explore the bulk-boundary
correspondence in the physical state by introducing an entanglement cut in the spatially periodic
physical state.

\section{Entanglement} \label{secE}

\begin{figure}
\includegraphics[width=.6 \linewidth,keepaspectratio]{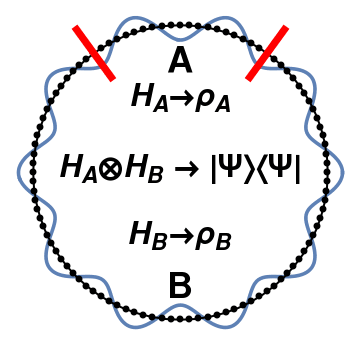}
\caption{The entanglement cut considered. The blue wavy line represents the pure state
for a spatially periodic (infinite) wire. We consider three kinds of pure states. One
is the ground state of the static Hamiltonian, the second is the FGS, and the third
is a physical state obtained from unitarily evolving the static ground state following a quench
of the periodic drive.
The density matrix of the full Hilbert space can be constructed
from the pure state, and the corresponding mixed states of the subsections indicated are found from
formally tracing out the complementary degrees of freedom.}
\label{entcut}
\end{figure}

We consider the entanglement cut shown in figure \ref{entcut}.
Our full state is a pure state which could be one of the following three with periodic boundary conditions.
One is the ground state of the static Hamiltonian, the second is the FGS, and the third is
the time evolved quenched state. The entanglement cut is in real space, and the reduced density matrix is obtained from tracing out
the complementary degrees of freedom.

Carrying out such a partial trace is not a simple task. However for quadratic Hamiltonians, there exists a relation between the reduced density
matrix and a matrix of single particle correlations on the section of the lattice of interest
\cite{Eisler2009}. By Wick's theorem, all free-fermion cumulants will factor, including those that occur
within the entanglement cut. Since our reduced density matrix must recreate this decomposition, the reduced density matrix must have the following form,
\begin{align*}
\rho \propto e^{-\mathcal{H}},
\end{align*}
where $\mathcal{H}$ is quadratic in the fermionic operators.
To derive the ES which is the eigenvalues of $\mathcal{H}$, we will resort to the Majorana basis description\cite{Canovi14}.

Our ``complex" fermions in the region of interest (say A) are broken into the Majorana basis much like real and imaginary
parts of a complex number,
\begin{align}
	c_i &= \frac{1}{2} \left( a_{2i-1} + i a_{2i} \right), \\
	c_i^\dagger &= \frac{1}{2}\left(a_{2i-1} - i a_{2i} \right).
\end{align}
We have $\{a_{2i},a_{2j}\} = \{a_{2i-1}, a_{2j-1}\} =  2 \delta_{i,j}$ and $\{a_{2i},a_{2j-1}\} = 0$.

In order to study the Majorana correlation matrix,
\begin{align*}
	\tilde{C}_{n,m} &= \Tr\left[ \rho a_n a_m \right] = \langle \Psi(t)|a_n a_m |\Psi(t)\rangle,
\end{align*}
it is convenient to
define $C_{i,j} $=$ \Tr\left[ \rho c_i^\dagger c_j \right] $=$ \bra{\Psi(t)} c_i^\dagger c_j \ket{\Psi(t)}$ and
$F_{i,j} $=$ \Tr \left[ \rho c_i^\dagger c_j^\dagger \right] $=$ \bra{\Psi(t)}c_i^\dagger c_j^\dagger \ket{\Psi(t)}$, so that,
\begin{align*}
	\tilde{C}_{2i-1,2j-1}
	&= \delta_{i,j} + 2 i \Im\left[C_{i,j} +  F_{i,j}\right],\\
	\tilde{C}_{2i-1,2j}
	&= i \delta_{i,j} - 2 i \Re \left[C_{i,j} - F_{i,j} \right],\\
	\tilde{C}_{2i,2j-1}
	&= -i \delta_{i,j} + 2 i \Re \left[C_{i,j} + F_{i,j} \right],\\
	\tilde{C}_{2i,2j}
	&= \delta_{i,j} + 2i \Im \left[C_{i,j} - F_{i,j} \right].
\end{align*}
We group together the neighboring matrix elements into a single matrix,
\begin{align} \label{corr}
	\mathbb{C}_{i,j} &= \left(\tilde{C} - 1\right)_{i,j}\notag \\
	&= i
	\begin{pmatrix}
		2\Im\left[ C_{i,j} +  F_{i,j}\right] & \delta_{i,j} - 2 \Re \left[C_{i,j}- F_{i,j} \right]\\
		 -\delta_{i,j} +2 \Re \left[C_{i,j} + F_{i,j} \right]& 2 \Im\left[C_{i,j} - F_{i,j} \right]
	\end{pmatrix},
\end{align}
where $i,j$ index the physical sites within the entanglement cut.
$\mathbb{C}$ is hermitian and purely imaginary.

When the system is the static ground state, we can assume that the $C,F$ are purely real.
In finding the spectrum  of the static system we can rearrange the rows and columns inside the determinant to bring it to the form,
\begin{equation*}
	\det\left[\mathbb{C}- \lambda\right] =  \det\begin{pmatrix}
		 - \lambda &  i -2 i \left[ C -  F \right]\\
		-i +  2 i \left[ C +  F \right] &  -\lambda
	\end{pmatrix}.
\end{equation*}
Using the Schur complement
$$\det\begin{pmatrix} A & B \\ C & D \end{pmatrix} = \det D \det \left( A - B D^{-1} C \right), $$ we
arrive at the following characteristic equation,
\begin{equation*}
\det \left( \lambda^2 - \left(1 - 2 \left( C + F\right) \right) \left(1 - 2 \left(C - F\right) \right)\right) = 0,
\end{equation*}
which is the same as the eigenvalue equation found in \cite{Lieb61,Eisler2009}, and was shown to give the
static entanglement spectra. We now proceed to show that the more general Majorana correlation matrix in equation~\eqref{corr}
will give the ES in the time dependent case.

A general quadratic Hamiltonian, such as our entanglement Hamiltonian, can be written in terms of Majorana fermions as,
\begin{equation}
	\mathcal{H} = i \sum_{m,n} w_{m,n} a_m a_n.
\end{equation}
Where $w$ is real and anti-symmetric; this means that there exists an orthogonal transformation
that will bring $\mathcal{H}$ into a block diagonal form,
\begin{equation*}
	a_n = \sum_m O_{n,m} \gamma_m.
\end{equation*}
The transformation being orthogonal is important as our new operators are
still Majorana fermions $\gamma^\dagger  = \gamma $. This transformation will
bring the entanglement Hamiltonian to the following form,
\begin{align*}
	\mathcal{H} &= i \vec{a}\cdot w\cdot \vec{a} = i \vec{\gamma}\cdot O^T\cdot w\cdot O\cdot \vec{\gamma},\\
	  &= \frac{i}{4}\vec{\gamma} \cdot \left( \sum_i \varepsilon_i i \sigma_y \right) \cdot \vec{\gamma},\\
	  &= \frac{i}{2}\sum_i \varepsilon_i \gamma_{2i-1} \gamma_{2i}.
\end{align*}
The Pauli matrix acts on the $2i-1, 2i$ sub-basis for each $i$.
As always we are free to add a constant energy to our Hamiltonian,
\begin{equation}
	\mathcal{H} = \frac{1}{2}\sum_i \varepsilon_i \left(1+ i \gamma_{2i-1} \gamma_{2i} \right).
\end{equation}
This is the same form if we had used a Bogoliubov transformation on the complex fermions and then
performed the transformation to the Majorana basis afterwards.

In terms of $\gamma$, the reduced density matrix has the diagonal form,
\begin{align*}
\rho = \prod_i \left[\frac{e^{-\frac{\varepsilon_i}{2}
\left( 1 + i \gamma_{2i-1} \gamma_{2i} \right) }}{1 + e^{-\varepsilon_i}} \right].
\end{align*}
We now insert this form into our correlation matrix definition,
\begin{align*}
	\tilde{C}_{n,m}
&= \sum_{p,q} \Tr \left[
\prod_i \left[\frac{e^{-\frac{\varepsilon_i}{2}\left( 1 + i \gamma_{2i-1} \gamma_{2i} \right) }}
{1 + e^{-\varepsilon_i}} \right]O_{n,p} O_{m,q}\gamma_p \gamma_q \right].
\end{align*}
The only terms that survive the trace are when $p = q$,
when $p = 2j$, $q = 2j -1$ and when $p = 2j -1$, $q = 2j$.
The operator $i \gamma_{2j-1} \gamma_{2j}$ will measure the fermion parity at the site $j$, thus we perform the trace
and arranging the sum into the even-odd matrix notation as before,
\begin{align*}
	\tilde{C}_{n,m}
	&=  \sum_{j}O_{n,j}
\begin{pmatrix}
1 & -i \tanh\left( \frac{\varepsilon_j}{2} \right)\\
i\tanh\left( \frac{\varepsilon_j}{2} \right)& 1
\end{pmatrix} O^T_{j,m}.
\end{align*}
The orthogonal transformation that was performed on the entanglement Hamiltonian also
block-diagonalizes $\tilde{C} -1$.
\begin{equation*}
	O^T\tilde{C}O = \tilde{C}' = \sum_{i} \left[1 + \sigma_y \tanh\left(\frac{\varepsilon_i}{2} \right) \right],
\end{equation*}
so that spectrum of $\tilde{C} - 1$ will yield the ES.

Defining $E = \tanh \frac{\varepsilon_k}{2} $ our spectrum will lie between $-1, 1$ and
the entanglement entropy (EE) will take the form,
\begin{equation} \label{ent}
 S = -\sum_k \left[
\frac{1+E}{2}\log \left(\frac{1+E}{2} \right)+\frac{1-E}{2}\log\left(\frac{1-E}{2}\right)\right].
\end{equation}
A value of $E = 0$ corresponds to a maximally entangled Schmidt state, whereas $E = \pm 1$
corresponds to a state that is minimally entangled.

In what follows we will discuss only the ES and not the EE. This is because~\cite{Yates16},
the EE of the FGS behaves generically as that of a ground state wavefunction of a gapped Hamiltonian by showing
area law scaling. The quenched or physical state on the other hand, at long times after the quench, shows
a saturation to a volume law scaling for the EE.
The volume law reflects the finite density of excitations that are always present when the periodic drive
is resonant, and is again a behavior generic to excited states. Since the EE does not show any features of topology,
in the remaining paper we will study the ES alone.

\section{Entanglement spectra of the static ground state}\label{secG}

For the static case, the $C,F$ matrices are purely real, and take the form,
\begin{align}
C_{i,j}
&= \frac{\delta_{i,j}}{2} + \int_{0}^\pi \frac{dk}{2 \pi}
\frac{ \cos (k(i-j))\left(\frac{\mu_0}{2} + t_h \cos k\right)}{\sqrt{\left( \frac{\mu_0}{2} +  t_h \cos k\right)^2 + \Delta^2 \sin^2 k}},\\
F_{i,j}
&= -\int_0^\pi \frac{dk}{2 \pi} \frac{\Delta \sin (k(i-j)) \sin k}
	{\sqrt{\left(\frac{\mu_0}{2} +  t_h \cos k \right)^2 +  \Delta^2 \sin^2(k)}}.
\end{align}
Inserting these relations into our Majorana correlation matrix \eqref{corr}, yields the ES
for the static ground state upon diagonalization.

\begin{figure}
\includegraphics[width=.95\linewidth]{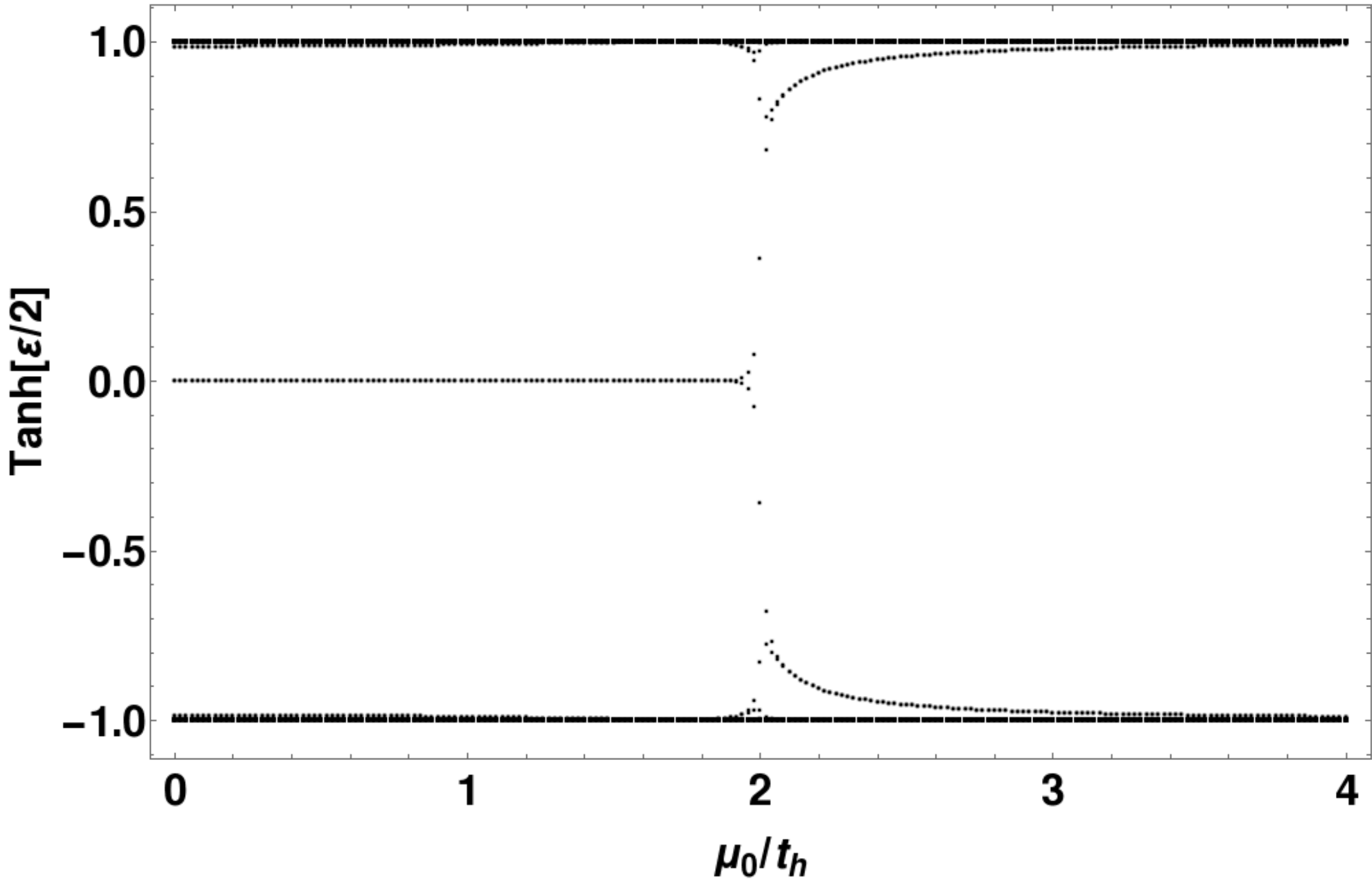}\\
\caption{
ES verses $\mu_0$ of the static ground state for an entanglement cut of 50 sites.
Here $\Delta/t_h = .5$. $0<\mu_0/t_h<2$ corresponds to the MZM phase, $2<\mu_0/t_h<4$
corresponds to the trivial phase.
The ES
appears as a spectrally flattened version of the physical edge energy spectrum. The
topologically protect edge states reside at zero entanglement energy.
}
\label{ent_gs}
\end{figure}

\begin{figure}
\includegraphics[width = .95\linewidth,keepaspectratio]{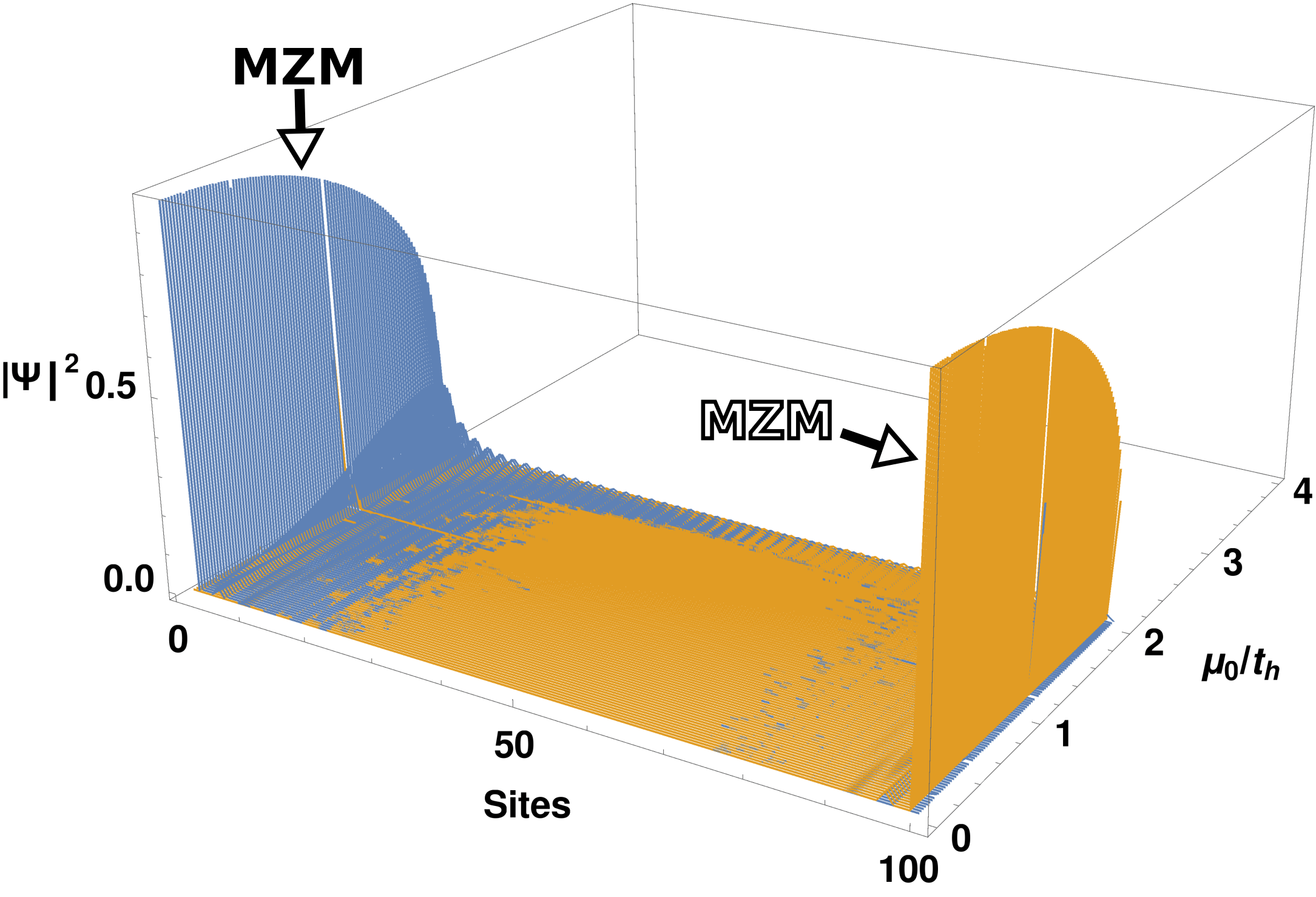}
\caption{
The Schmidt states for an entanglement cut of 50 sites in the static ground state, $\Delta/t_h = .5$.
Plotted are the Schmidt states within a window of $|.4|$ to an entanglement energy of $0$.
Distinct colors (red, orange, green, blue) denote distinct eigenstates.
We note that the Schmidt states
have double the number of sites because these states are in the Majorana basis. This
doubling is equivalent to the effective spinor degree of freedom describing the superconductor.
}
\label{ent_gs_edges}
\end{figure}

Figure \ref{ent_gs} shows the ES of the ground state.  Comparing to the off-resonant section of figure \ref{finite_om5}, the ES
correctly recreates the energy level spectrum of the Kitaev chain with physical boundaries, but with the bands flattened\cite{Fidkowski10}.
We also note that figure \ref{ent_gs} has a large ``entanglement gap" which will remain open in the absence of bulk excitations\cite{Yates16}.
Figure \ref{ent_gs_edges} shows that the zero energy states are edge states.

All levels away from the singular points of the topological phase transition show a double degeneracy. This
corresponds to the inversion operation with respect to the center of the entanglement cut.
The regions close to the transition correspond to diverging length scales and hence the anti-symmetric and symmetric states separate.

Figure \ref{ent_gs_edges} shows the Schmidt states within
a window $|.4|$ of zero entanglement energy $\tanh(\epsilon/2)$. This window reliably captures the topological Majorana zero modes, and is
employed for all plots showing Schmidt states.
Note that even the trivial phase could host edge-modes, but these edge modes are non-topological in that they
are composed of complex or Dirac fermions rather than Majorana fermions, and are located at an entanglement energy closer to the bulk entanglement
energy $\pm 1$ (and therefore do not appear in Figure \ref{ent_gs_edges}). Weak perturbations such as disorder,
not considered in this paper, will not protect such
complex fermions from merging with the bulk.

To understand what states will be protected in a more general system,
we must investigate the effect of allowed, non-interacting couplings.
Since the entanglement Hamiltonian is closely related to a band-flattened version of the parent Floquet Hamiltonian
if placed in finite geometry, we consider the robustness of edge states in the ES to perturbations which preserve the
anti-unitary symmetries of PHS and TRS.

PHS dictates that the positive entanglement energy states are related to the
negative entanglement energy states through particle-hole conjugation. Since under particle-hole conjugation,
a Majorana state transforms to itself, $\epsilon = 0, \pm\Omega/2$ are
the only viable energy levels for Majorana states in the quasi-energy spectrum.
In the entanglement Hamiltonian, only $\epsilon = 0$ is possible. Thus any coupling between Majorana states and a
bulk state away from $\epsilon = 0$ is not allowed. If such a coupling did exist and
successfully gapped out the Majorana state, then the newly gapped state would also have
to be described by a complex fermion, but there are no extra Majorana operators for the
coupled Majorana to join together with and form a complete fermion.
Thus PHS symmetry preserving couplings cannot move the zero energy Majorana level.

Now we discuss stability with respect to TRS preserving couplings.
Breaking a fermion into its Majorana operators $c_i \propto a_i + i b_i$
($a/b \leftrightarrow$ odd/even sites),
TRS is the pair of statements, $\mathcal{T}a_i \mathcal{T}^{-1} = a_i $ and
$\mathcal{T} b_i \mathcal{T}^{-1} = - b_i $. Thus, if we are to have a TRS entanglement
Hamiltonian, couplings of the form $i a_i a_j$ and $i b_i b_j$ are prohibited. Since a Majorana mode
corresponds to an $a$-type fermion on one edge, and a $b$-type fermion on the other, local
couplings such as $i a_i b_j$, while TR preserving, cannot affect the Majorana mode.

Thus in summary, for the static ground state ES, the topological Majorana edge states are
protected by a large gap, PHS and TRS. In the trivial phase, the edge states are complex fermions
and thus are sensitive to simple perturbations that couple to their occupation.

\subsection{Analytic solution for edge states in the ES}
We would like to understand how the edge states are created in the ES analytically. This
can be done at some special points and mirror the reasoning provided in Kitaev's original paper ~\cite{Kitaev01}, but for the
entanglement cut.

We set $\Delta/t_h = 1$ and probe the trivial region by letting $\mu_0\rightarrow \infty$, and the topological region by setting $\mu_0=0$. In the
trivial region ($\mu_0\rightarrow \infty$) the two correlators become,
\begin{align*}
C_{i,j} &\approx \delta_{i,j},\\
F_{i,j} &\rightarrow0.
\end{align*}
Plugging this into our Majorana correlator, \eqref{corr},
\begin{align*}
	\mathbb{C}_{i,j}
	&= \begin{pmatrix}
		0&  -i \delta_{i,j}\\
		i \delta_{i,j}& 0
	\end{pmatrix}.
\end{align*}
Our Majorana correlator becomes block-diagonal with degenerate bands at $\pm 1$, thus reproducing the ES of the trivial phase.

Now we consider the topological phase ($\mu_0 = 0$),
\begin{align*}
C_{i,j} &= \frac{\delta_{i,j}}{2} + \int_{0}^\pi \frac{dk}{2 \pi}\cos (k(i-j))\cos k,\\
F_{i,j} &= -\int_0^\pi \frac{dk}{2 \pi} \sin (k(i-j)) \sin k.
\end{align*}
The above implies,
\begin{align*}
	\left[C + F\right]_{i,j} &= \frac{\delta_{i,j}}{2} + \frac{\delta_{i-j,-1}}{2},\\
	\left[C - F\right]_{i,j} &= \frac{\delta_{i,j}}{2} + \frac{\delta_{i-j,1}}{2},
\end{align*}
which when inserted into \eqref{corr}, gives,
\begin{align*}
	\mathbb{C}_{i,j} 
	&=
	\begin{pmatrix}
		0&-i \delta_{i-j,1}\\
		i\delta_{i-j,-1}&0
	\end{pmatrix}.
\end{align*}
This corresponds to the same block-diagonal matrix as for the trivial case
with the modification of an empty top and bottom row.
It is the same matrix in the bulk with degenerate bands at $\pm 1$, but 
we now have null vectors that occupy the first and last sites. These null vectors
denote the Majorana edge modes in the ES.
Thus we have re-derived the Kitaev picture, but now for our entanglement cut.

\section{Entanglement spectra of FGS}\label{secF}
We now wish to calculate the correlators as a function of time.
We can construct the $C,F$ matrices from our knowledge of the full time-evolved wave function.
Denoting the Floquet ground and excited states as $\downarrow$ and $\uparrow$ respectively,
such that $\ket{\text{FGS}(t)}_k = \alpha_\downarrow(k,t) c_k^\dagger c_{-k}^\dagger \gs + \beta_\downarrow(k,t) \gs$,
we find,
\begin{align}
C_{i,j}
&= \frac{1}{\pi}\int_0^\pi dk \cos(k(i-j)) \left[
|\rho_{\downarrow} \alpha_\downarrow|^2 +\right.\notag \\
&\left.\qquad |\rho_{\uparrow} \alpha_\uparrow|^2
+ e^{i \epsilon_k t}\rho_{\downarrow} \rho_{\uparrow}^*\alpha_{\downarrow}\alpha_{\uparrow}^*
+ e^{-i \epsilon_k t}\rho_\uparrow \rho_\downarrow^*\alpha_\uparrow \alpha_\downarrow^*
\right],
\label{c_driven}\\
F_{i,j}
&= \frac{i}{\pi} \int_0^\pi dk \sin(k(i-j)) \left[
|\rho_\downarrow|^2 \beta_\downarrow \alpha_\downarrow^* + \right.\notag\\
&\left. \qquad |\rho_\uparrow|^2 \beta_\uparrow \alpha_\uparrow^*
+ e^{i \epsilon_k t}\rho_\downarrow \rho_\uparrow^*\beta_\downarrow \alpha_\uparrow^*
+ e^{-i \epsilon_k t}\rho_\uparrow \rho_\downarrow^*\beta_\uparrow \alpha_\downarrow^*
\right].
\label{f_driven}
\end{align}

We are interested in two driven states, the FGS, and a physical state obtained from a quench. The physical
state will correspond to utilizing the full expressions for $C$ and $F$ with $\rho_{\uparrow},\rho_{\downarrow}$
given in Eq.~\eqref{rhodef}. The FGS ES will
be determined from the above after setting $\rho_\downarrow = 1$ and $\rho_\uparrow = 0$
for all $k$. The quench state discussion will follow the FGS in section~\ref{secQ}.

Unlike the quasi-energy spectrum, the ES is not periodic, and has only one gap. Any topological
edge modes have to lie within this gap. In this sense, the ES of the FGS creates an edge spectrum with only
``half" the information compared to the quasi-energy spectrum at physical boundaries.
The loss of ``half" the information will result in the MZM and MPM states both having the
same entanglement energy, $\epsilon = 0$.

\begin{figure}
\includegraphics[width = .95\linewidth,keepaspectratio]{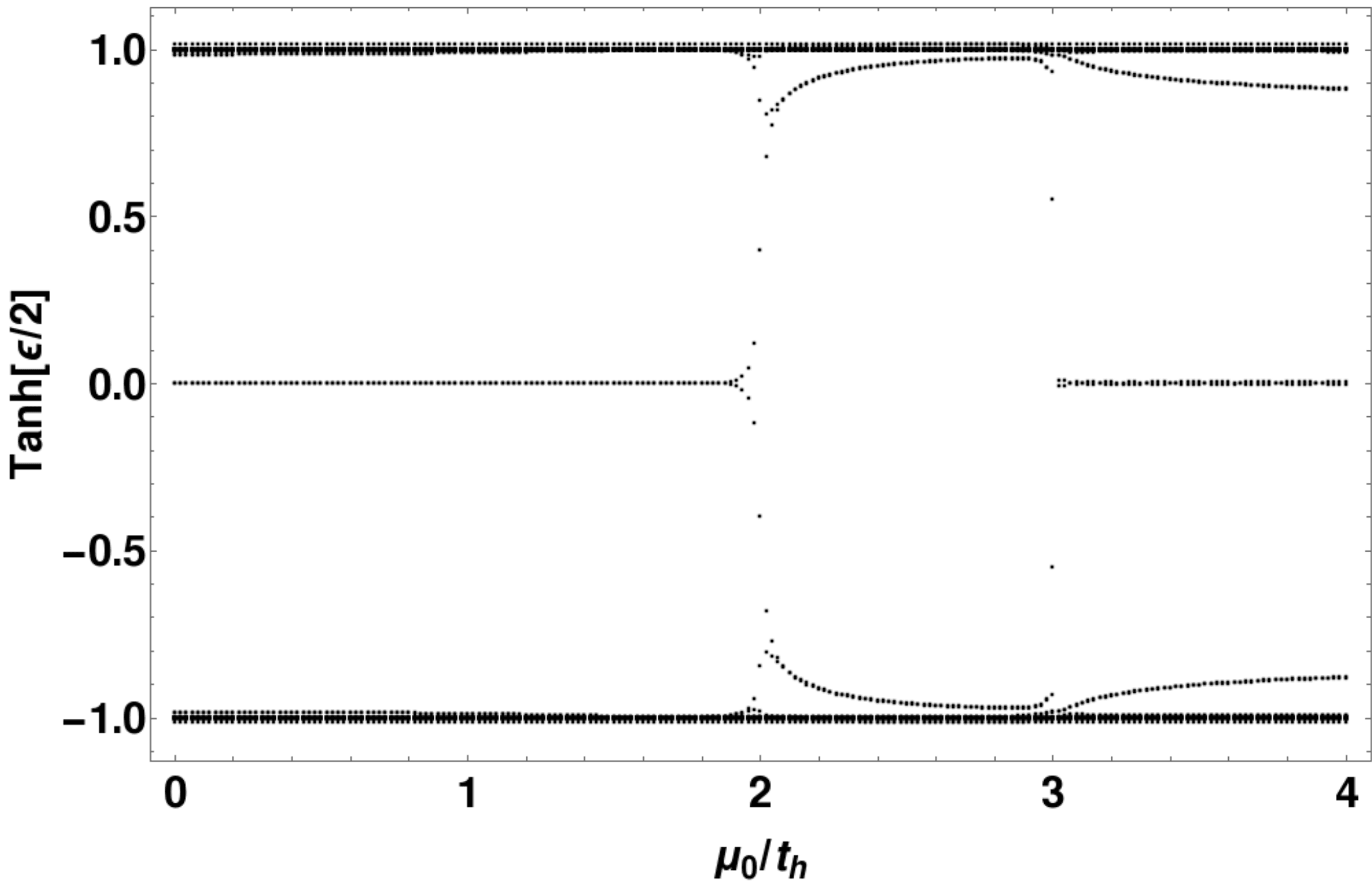}
\caption{
ES for the FGS at the start of a period, $\Omega/t_h = 5$, $\Delta/t_h = .5$, $\xi/t_h = 2.$, and $N = 35$.
$0<\mu_0/t_h<2$ corresponds to the MZM phase, $2<\mu_0/t_h<3$ corresponds to the trivial phase, and
$3<\mu_0/t_h<4$ corresponds to the MPM phase.
The spectrum here appears as a spectrally flattened version of the quasi-energy levels of the physical wire,
where the MPM states are moved from $\epsilon = \pm \Omega/2 \rightarrow 0$.
The ES here is not sensitive to the chosen time during the drive. There is very little
motion from the ``bulk" excitations (at $\pm 1$), while the ``low-energy" states (at $0$) remain fixed.
}
\label{om5f}
\end{figure}

\begin{figure}
\includegraphics[width = .95\linewidth,keepaspectratio]{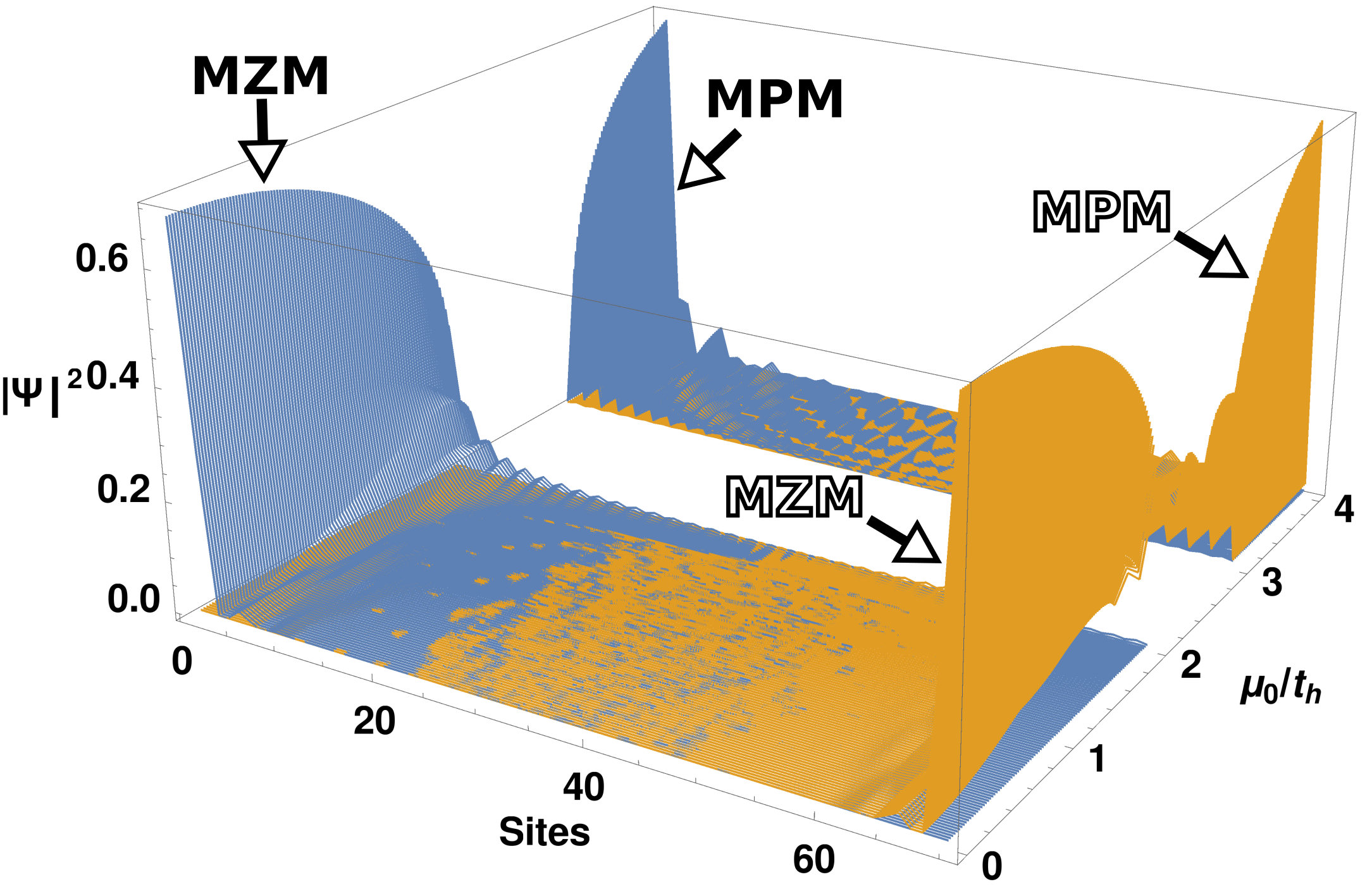}
\caption{ES edge Schmidt states for the FGS at the start of a period, $\Omega/t_h = 5$, $\Delta/t_h = .5$, $\xi/t_h = 2$, and
$N = 35$. The states at the center of the ES for the FGS correspond to Schmidt states located at either ends of the wire,
for both the MZMs and MPMs.
Distinct colors (red, orange, green, blue) denote distinct eigenstates.
}
\label{om5fe}
\end{figure}

\begin{figure}[t]
\includegraphics[width = .95\linewidth,keepaspectratio]{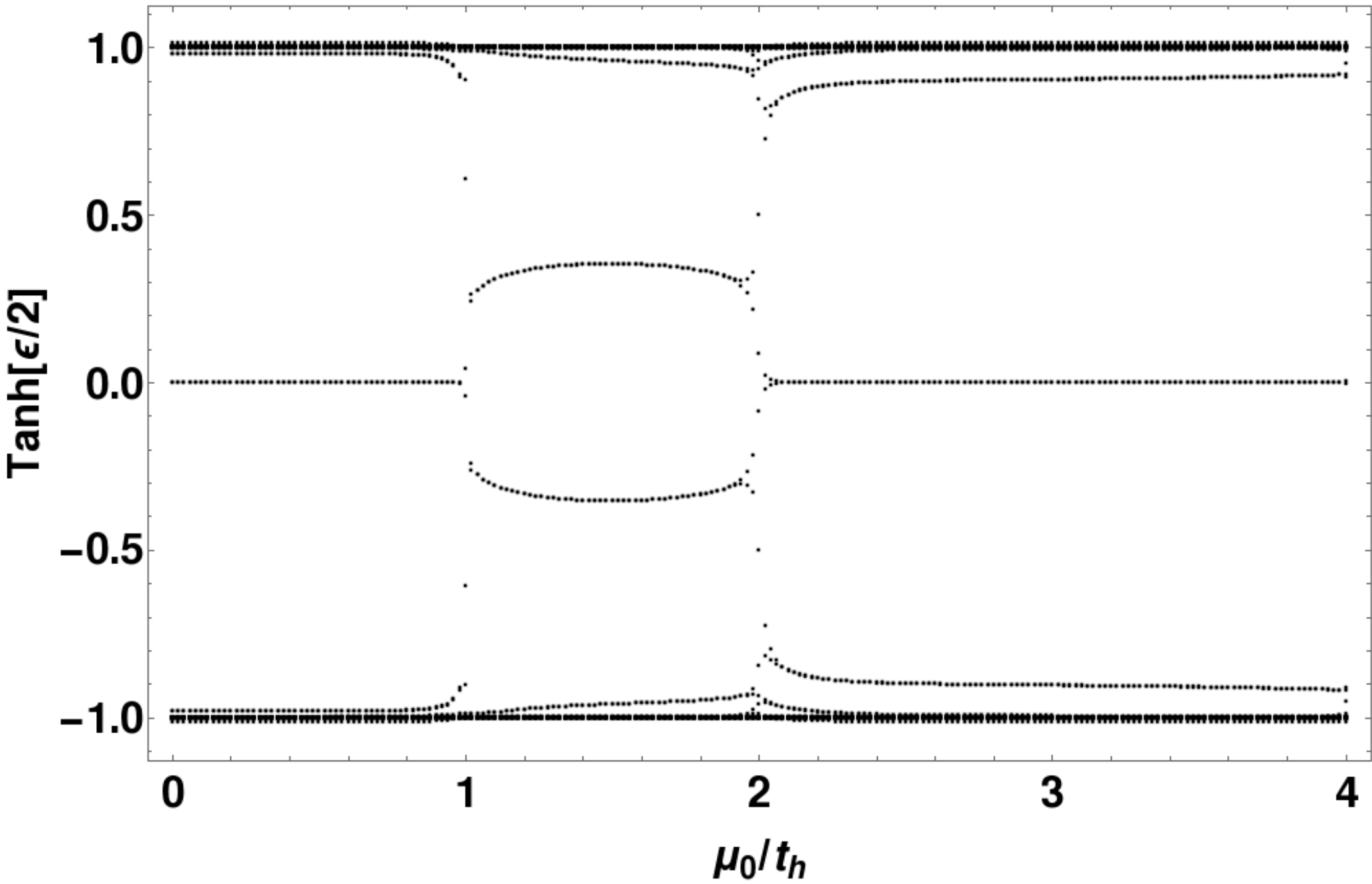}
\caption{
ES for the FGS at the start of a period, $\Omega/t_h = 3$, $\Delta/t_h = .5$, $\xi/t_h = 2$, and $N = 35$.
$0<\mu_0/t_h<1$ corresponds to the MZM phase, $1<\mu_0/t_h<2$ corresponds to the MZM\&MPM phase, and
$2<\mu_0/t_h<4$ corresponds to the MPM phase.
The ES reproduces the edge states for the MZM phase and the MPM phase at zero entanglement energy.
The MZM\&MPM states are gapped out in the ES at the chosen time. See figure \ref{om3ft}
for their behavior at other times.
}
\label{om3f}
\end{figure}

\begin{figure}
\includegraphics[width = .95\linewidth,keepaspectratio]{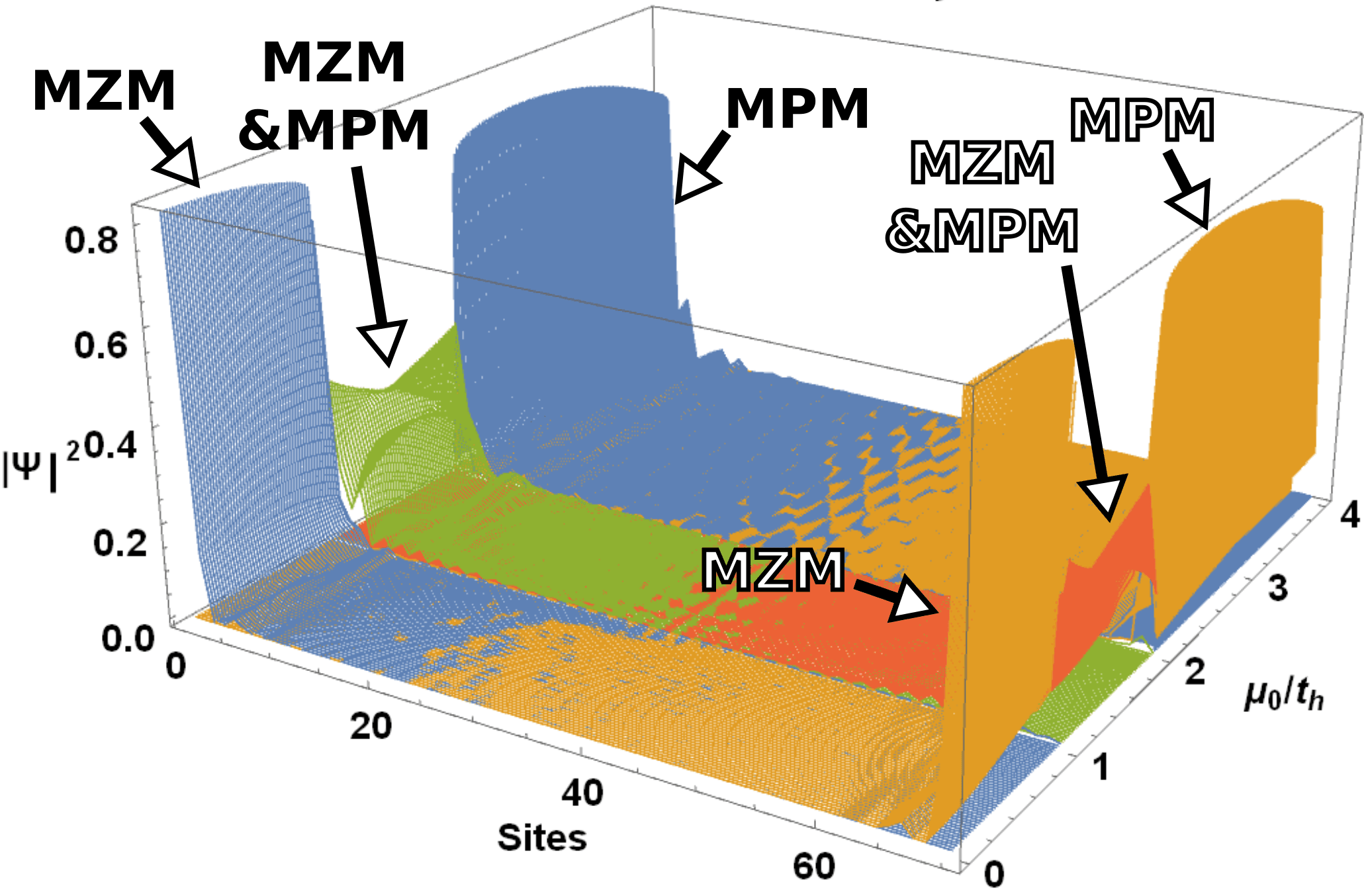}
\caption{ES edge Schmidt states for the FGS at the start of a period, $\Omega/t_h = 3$, $\Delta/t_h = .5$, $\xi/t_h = 2$, and
$N = 35$. The states at the center of the ES for the FGS correspond to Schmidt states located at either ends of the cut,
for both the MZMs and MPMs.
Distinct colors (red, orange, green, blue) denote distinct eigenstates.
The Schmidt states corresponding to the low-lying states in the ES for the MZM\&MPM phase (4 total, with 2 on each edge)
are still edge states, but with longer localization lengths.
}
\label{om3fe}
\end{figure}

\begin{figure}
\includegraphics[width = .95\linewidth,keepaspectratio]{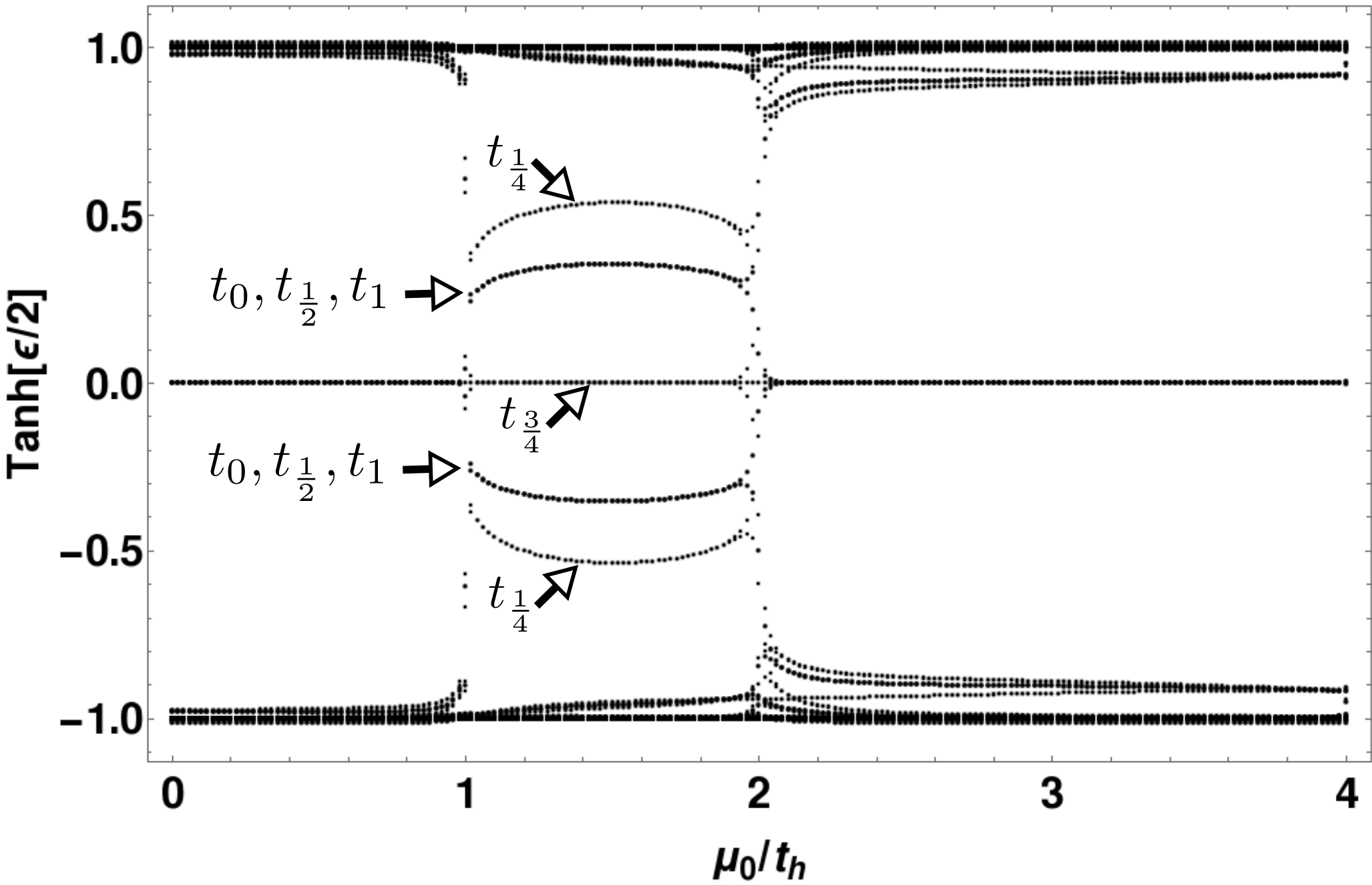}
\caption{
ES for the FGS at four different times $t=$ ($0,T/4,T/2,3T/4,T$) for
$\Omega/t_h = 3$, $\Delta/t_h = .5$, $\xi/t_h = 2.$, and $N = 35$. In the MZM phase ($0<\mu_0/t_h<1$) and the MPM phase
($2<\mu_0/t_h$), the ES is essentially unchanged over the period.  In the MZM\&MPM phase ($1<\mu_0/t_h<2$), the edge modes
oscillate dramatically during the drive, with
$\mathbb{Z}_0+\mathbb{Z}_{\pi}=2$ Majorana edge modes at $t^*=3T/4$, and $|\mathbb{Z}_0-\mathbb{Z}_{\pi}|=0$ Majorana edge modes at all other times.}
\label{om3ft}
\end{figure}

\paragraph*{\textbf{Discussion of the FGS plots.}}
We will now make a series of observations from the plots for the ES of the FGS, but will explain these observations
in section~\ref{secS}. In particular, since the FGS ES is
constructed from the pure FGS alone, we will find it convenient to explain some of the unusual features in the ES through a
spinor description of the FGS in section~\ref{secS}.

We first focus on the FGS ES in figures \ref{om5f} and \ref{om5fe} for  $\Omega/t_h = 5$, which contain the three phases
MZM, trivial, and MPM. A comparison with figures \ref{finite_om5}  and \ref{finite_om5_edges}
shows that the ES correctly reproduces the topological edge states of the physical edges.
The ES is a spectrally flattened version of the quasi-energy spectrum for the
physical edges, with the important modification that the $\pi$ modes now reside at zero entanglement
energy. Further, the region of the ES that corresponds to off-resonant drives closely resembles the ES of the static
ground state and the physics of this portion of the ES is similar to the static case.
In addition, the MPM and MZM phases behave in a largely similar manner. Figure \ref{om5fe} shows the exponential 
localization of both the MPM and MZM modes, within a few lattice sites of the edges. 
The decay length is governed by the entanglement gap in the ES, which for figure \ref{om5f} is nearly constant
throughout the period.

We now shift our focus to figures \ref{om3f} and \ref{om3fe}, where $\Omega/t_h = 3$ and
corresponds to a resonance occurring while still in the MZM phase. Here we have a MZM,
MZM\&MPM, and MPM phase for the range of $\mu_0$ shown.
We will now discuss the new MZM\&MPM phase, as the remaining phases are largely the
same as that of the example discussed above for $\Omega/t_h = 5$.

The MZM\&MPM phase is interesting because the MZM and MPM states gap each other out in the ES.
This is in contrast to figures \ref{finite_om3} and \ref{finite_om3_edges} where
the MZM\&MPM phase is simply the combination of the individual phases, without any coupling between them, as predicted by the
$\mathbb{Z} \times \mathbb{Z}$ index. We find that $\mathbb{Z}\times\mathbb{Z}$,
no longer holds in the ES as is evident by the gap opening. By studying the Schmidt states we note that the gap opening
{\bf does not} involve the MZM and MPM modes merging with the bulk. Rather, we still have edge modes, but the nature of
the edge modes are different when MZM and MPM modes couple to each other.

Figure \ref{om3ft} shows the time dependence of the gap in the
MZM\&MPM phase. At most points during the drive, the phase is gapped, but at a
special point, the gap closes and both the MZM and the MPM reside at zero entanglement energy.
The time-dependence of the gap can be best understood through the spinor parameterization of the FGS
which we discuss in the next section.
Figure \ref{om3ft} also highlights the lack of time dependence in the ES for the topologically
protected states in the MZM and MPM phase, which as mentioned before behaves like
the ES of the static Hamiltonian. The ``bulk" states also show only a small amount of time dependence.

As far as the nature of the Schmidt states are concerned in the MZM\&MPM phase,
when the gap closes ($t=3T/4$ in figure~\ref{om3ft}), we have two Majorana modes
at each end of the cut. When the gap opens symmetrically around zero, this corresponds to a pair of edge modes
that are related by charge conjugation. Thus the positive entanglement energy edge mode is particle like, and the
negative entanglement energy edge mode is hole like.

This observed coupling between
$0$ and $\pi$ modes during a cycle leads to the conclusion that the $\mathbb{Z} \times \mathbb{Z}$ in the quasi-energy spectrum
becomes $|\mathbb{Z}_0 - \mathbb{Z}_\pi| \times |\mathbb{Z}_0 + \mathbb{Z}_\pi|$ in the ES, where the two integers now denote the
number of Majorana zero modes in the ES at the two TRS points $t^*=T/4,3T/4$.
We strengthen this observation further in the next subsection \ref{NNNsec} where
we generate more Majorana modes by introducing longer ranged hopping.


\subsection{Next-Nearest-Neighbor hopping} \label{NNNsec}
\begin{figure}
\includegraphics[width = .95\linewidth,keepaspectratio]{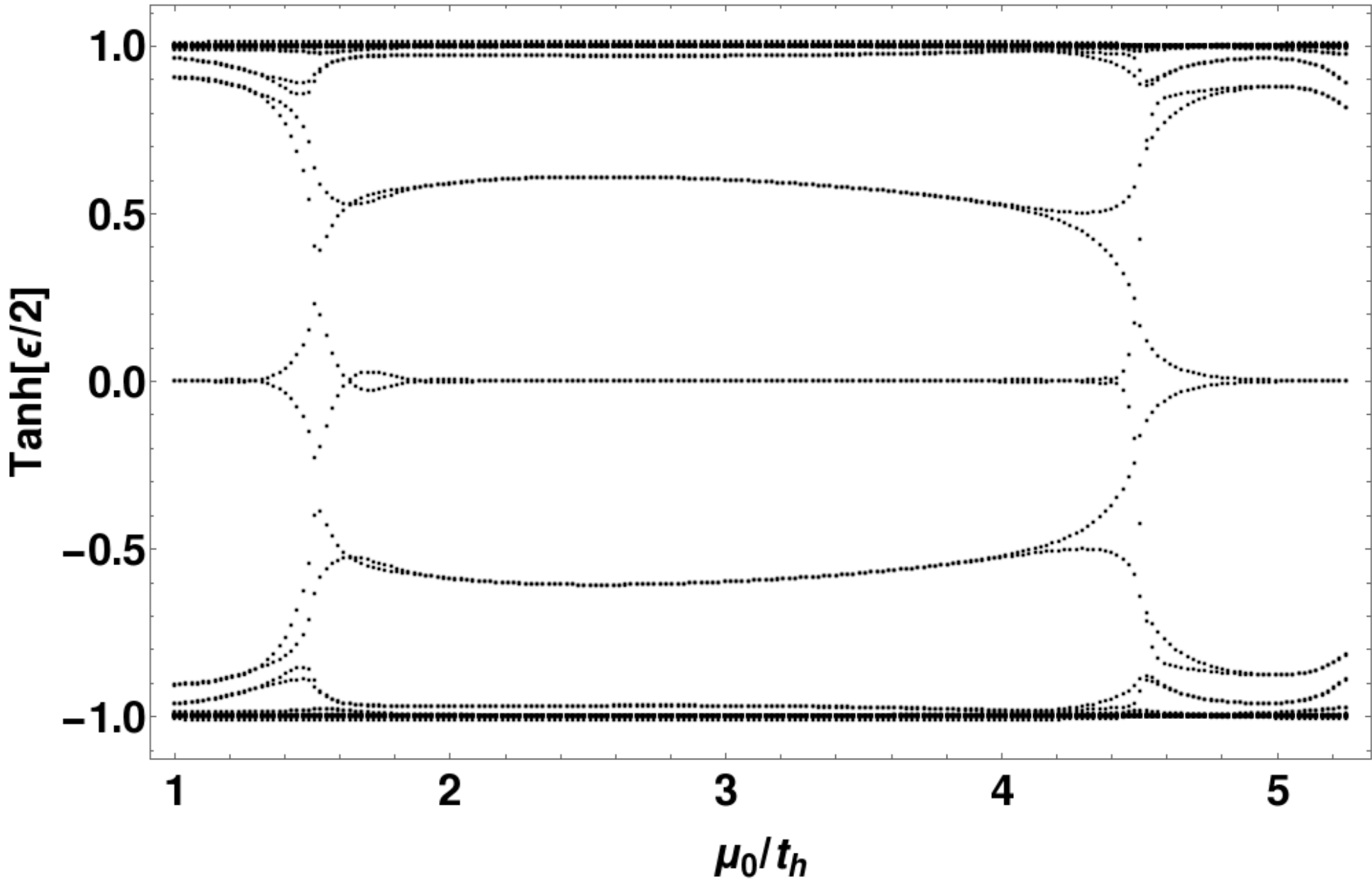}
\caption{
The ES of the NNN model, $\Delta = t_h$, $\Delta' = t_h'$, $\Omega/t_h = 4$,
$\xi/t_h  = 4$, $t_h'/t_h = -1.25$, $N = 35$, at the start of a period.
From left to
right, the phases are, 1MZM ($1<\mu_0/t_h<1.5$), 1MZM\&2MPM ($2<\mu_0/t_h<4$), 2MPM ($4.5<\mu_0/t_h$). For the
2MPM and 1MZM phases, the edge-modes are pinned at zero entanglement energy at all times.
}
\label{nnnpic}
\end{figure}

With NN hopping, $\mathbb{Z}_0,\mathbb{Z}_{\pi}$ take values $0,1$ and cannot therefore
differentiate between classes BDI and D.
In order to lift this ambiguity,
we would like to generate more MZMs and MPMs. The easiest way of creating more topologically protected edge modes is to introduce NNN hopping.
Here we use the model studied in \cite{Niu12} and turn on the NNN parameters ($t_h', \Delta'$).
The main conclusion from this section of the paper is that the NNN hopping produces larger number of edge states in our system and the
ES correctly detects these
edge states. In particular, for the example in figure
\ref{nnnpic}, we have three different phases. For the phase corresponding to $\mu_0/t_h>4.5$, we have four $\pi$ modes
at zero entanglement energy at all times, of which two
sit on one end of the entanglement cut and the other two at the other end. This phase has $\mathbb{Z}_0 =0, \mathbb{Z}_\pi=2$.
The gaplessness at zero energy holds true at all times and reveals that the
ES preserves $|\mathbb{Z}_0 - \mathbb{Z}_\pi|=2$.

\begin{figure}
\includegraphics[width=.95\linewidth,keepaspectratio]{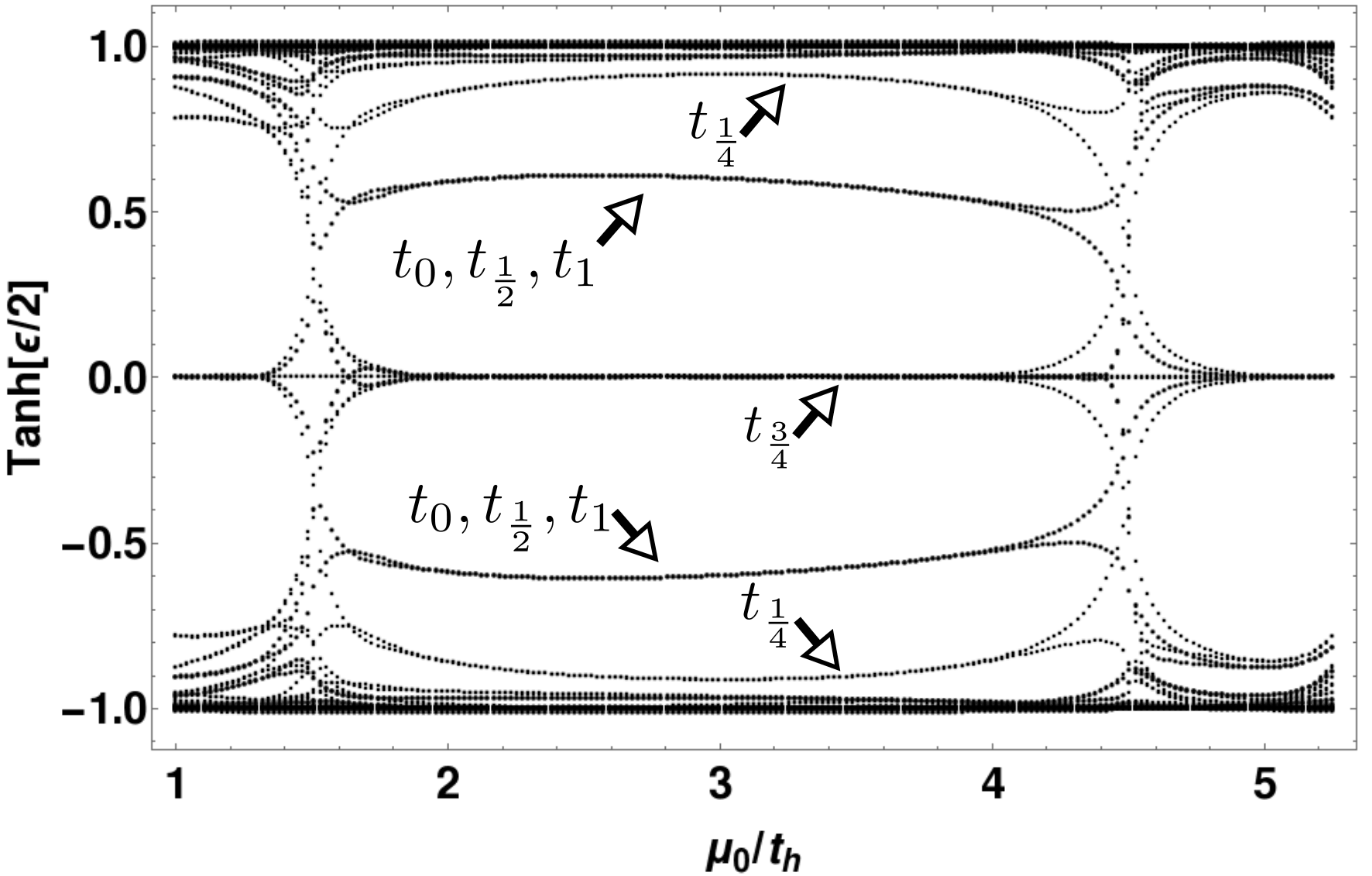}
\caption{
ES for the FGS with NNN terms,
$\Delta = t_h$, $\Delta' = t_h'$, $\Omega/t_h = 4$,
$\xi/t_h  = 4$, $t_h'/t_h = -1.25$, $N = 35$,
at four different times $t=$ ($0,T/4,T/2,3T/4,T$).
In the MZM phase ($0<\mu_0/t_h<1$) and the 2MPM phase
($4.5<\mu_0/t_h$), the ES is essentially unchanged over the period.  In the MZM\&2MPM phase ($2<\mu_0/t_h<4$), the edge modes
oscillate dramatically during the drive, with
$\mathbb{Z}_0+\mathbb{Z}_{\pi}=3$ Majorana edge modes at $t^*=3T/4$, and $|\mathbb{Z}_0-\mathbb{Z}_{\pi}|=1$ Majorana edge modes at all other times.
}
\label{nnnpic2}
\end{figure}

This point is further highlighted by studying the time-dependence of the gap. While the phases which contain only MZMs ($1<\mu_0/t_h<1.5$)
and only MPMs ($4.5<\mu_0/t_h$) show no significant time-dependence, in the central region ($2<\mu_0/t_h<4$) corresponding to
a phase containing both MZMs and MPMs, the time-dependence is dramatic. Comparing figure \ref{nnnpic} with figure \ref{nnnpic2}
shows that for this 2 MPMs\&1 MZM phase, a pair of levels remain at entanglement
energy of zero throughout the drive ($|\mathbb{Z}_0 - \mathbb{Z}_\pi|=1$). The gapped out states oscillate during the drive
and at a special point during the drive ($t^*=3 T/4$) also reside at zero entanglement energy, similar to the MZM\&MPM phase
in the NN hopping diagrams for $\Omega/t_h = 3$ (figure \ref{om3ft}).

Employing a spinor description in section~\ref{secS} we will explain the above observations, and also argue
that the reason why the the number of $|\mathbb{Z}_0 - \mathbb{Z}_\pi|$ Majorana modes
persist at other times besides the special TRS points is due to
our particular drive, namely one that couples to the chemical potential.
More generic TRS and PHS preserving periodic drives will give rise to additional couplings between the $|\mathbb{Z}_0 - \mathbb{Z}_\pi|$
Majorana modes away from $t^*$,
reducing the invariant in the ES during the period to $\mathbb{Z}_2$.

\section{Spinor description and topology}\label{secS}

We show that the FGS ES can be better understood through the spinor representation of the FGS.
We consider a spinor parameterized by the two angles $\alpha$ and $\beta$,
$\left( \cos \frac{\alpha}{2}, e^{i \beta} \sin \frac{\alpha}{2} \right)$
and consider the Bloch sphere parameterized by these angles. Consider the static
Hamiltonian for now. We know from PHS that
the $k= 0$ and $k = \pi$ points must lie at either the north or south poles, as the
symmetries force the $\sigma_{x,y}$ term in $H_{\rm BdG}$ to be odd in $k$. TRS on the other hand forces
the third Pauli matrix (in our case $\sigma_x$) to be absent so that the spinor lies in $y-z$ plane of
the Bloch sphere. This allows us to define a winding number for the number
of times the spinor winds around in a given plane
of the Bloch sphere.

NNN terms via $\Delta', t_h'\neq 0$,
does not change this, but includes the possibility of introducing another point $k^*$ different from $0$ and $\pi$,
where the spinor points either on the north or south pole. Thus with NNN terms, and TRS,
there is a possibility of introducing additional windings, and hence additional MZMs.

If the drive is resonant, then additional special points $k_{\pi}$ appear
where the spinors are constrained to be at the poles, but this constrain is true only at special TRS points of the drive.
For our drive, these special
TRS points during the cycle are $t^*=T/4,3T/4$. Thus at times $t^*$ we find well defined windings of
$|\mathbb{Z}_0-\mathbb{Z}_{\pi}|$ and $\mathbb{Z}_0 + \mathbb{Z}_{\pi}$ on the Bloch sphere.
Since our drive couples to the chemical potential, it has the additional property that the
spinors at $k=0,\pi, k^*$  stay pinned to the poles at all times.
This results in the $|\mathbb{Z}_0-\mathbb{Z}_{\pi}|$ winding
to be preserved at all other times besides the special $t^*$ points. A slightly different drive which coupled to $\Delta' \rightarrow \Delta' +
\delta \sin(\Omega t)$ would have the property that only the $k=0,\pi$ are constrained at the poles at all times. This would imply that
the ES will show $\mathbb{Z}_2$ invariance, with $|\mathbb{Z}_0-\mathbb{Z}_{\pi}|$ and $\mathbb{Z}_0 + \mathbb{Z}_{\pi}$ appearing only intermittently
in the ES at the two special $t^*$ times.

\begin{figure}
\includegraphics[width = .8\linewidth, keepaspectratio]{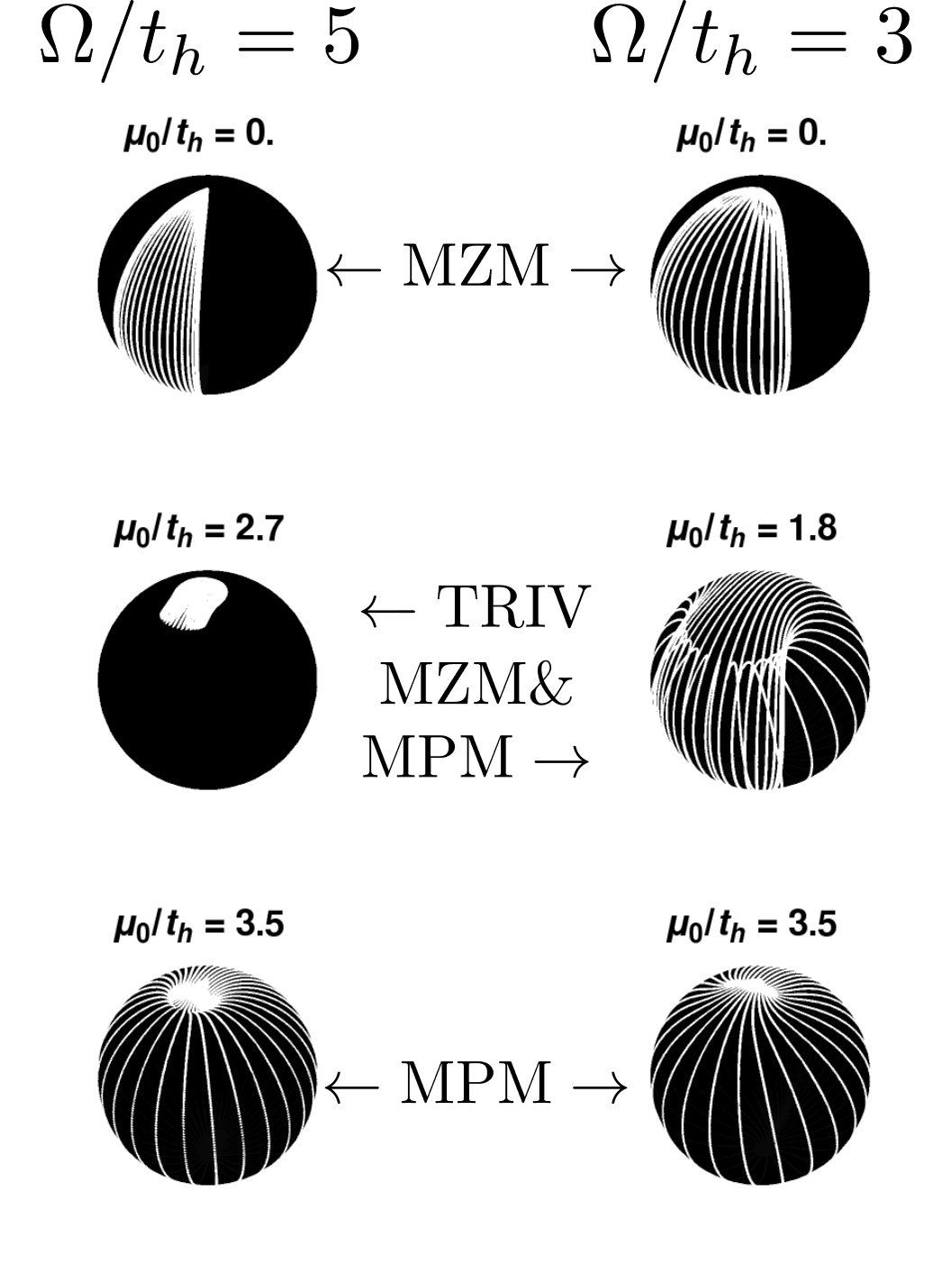}
\caption{
Spinor representation of FGS $\forall k \in [0,\pi], \forall t \in [0,T)$. The solid white lines
are for different times within a period.
Here both $\Omega/t_h = 3$ and $\Omega/t_h = 5$ are shown, with three choices of
$\mu_0/t_h$, each to highlight the generic behavior within the different phases. While both MZM and MPM straddle north
and south poles, the latter also covers the entire azimuthal angle $\beta \in [0,2\pi)$ during a drive cycle.
}
\label{spinors}
\end{figure}

We now discuss this basic picture with a specific example, with only NN hopping.
Figure \ref{spinors} shows how the FGS
for all $k$ wraps around the sphere for several times during the drive,
highlighting the generic behavior for the following phases: trivial, MZM, MPM, and MZM\&MPM.
The trivial phase fails to connect the north and south poles as expected. Since the drive is highly off-resonant in the trivial
phase, its only effect is to cause small deformations of the loop during the drive cycle.

In contrast to the trivial phase, the MZM, MPM phases always connect the two poles at all times.
The fact that these modes survive during the periodic drive is clear from the fact
that the drive still constrains the spinors to stay pinned at the poles at $k=0,\pi$.
The above picture explains why the ES shows static in time zero modes in the MZM and MPM phases.

The MPM phase, in addition to being pinned at the poles, has an associated wrapping in the $\beta$ angle as well.
The rotation in $\beta$ can be understood in the context of a rotating wave approximation.
The rotating frame effectively maps the resonant time-periodic
Hamiltonian (recall MPM phase is always associated with a resonance) into a Hamiltonian that appears like the static ground state in
the topological phase. Thus we regain the topological winding in the $\alpha$ parameter, but when we rotate
back into the lab frame, we acquire a relative phase between the two components of the spinor that is periodic with the
drive.

\begin{figure}
\includegraphics[width = .5\linewidth, keepaspectratio]{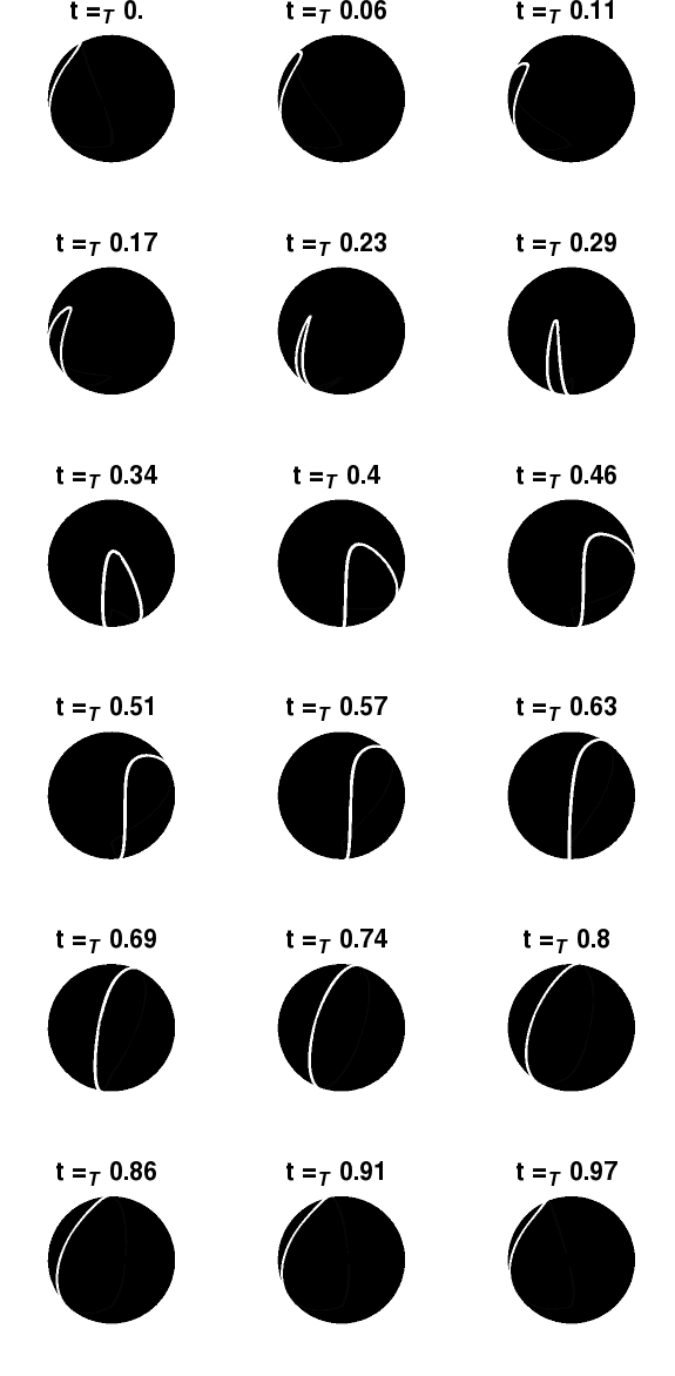}
\caption{
Spinor configurations of FGS $\forall k \in [0,\pi)$, for $\Omega/t_h = 3$, $\mu_0/t_h = 1.8$, which places
the FGS in the MZM\&MPM phase. This figure contains the same information as
figure \ref{spinors}, but now the time steps have been separated for their own
Bloch sphere to better highlight the lassoing effect that the combination of
MZM and MPM create. At a certain time during the cycle ($t= 3T/4$), the spinor straddles the north and south poles
and this coincides with appearance of zero modes in the ES in the MZM\&MPM phase (see figure \ref{om3ft}).
At other times, when the spinor configuration on the Bloch sphere is trivial,
the ES also does not show edge modes at zero entanglement energy.}
\label{spinors_alltimes}
\end{figure}

Now we turn to the case of the MZM\&MPM phase.
If we start in the MZM phase and increase $\mu_0$, the eventual phase transition and
introduction of the MPM, takes the end of the ``string" at the north pole and relocates it to the
south pole, thus ruining the non-trivial topology and gapping out the ES. However, focusing on figure \ref{spinors_alltimes},
one finds that at the special TRS point $t^*=3 T/4$, this string lassos around and straddles the two poles.

The spinors show nicely how the gap in the FGS ES changes with time.
The gap in the ES in the MZM\&MPM phase is sensitive  to the degree in which the string connects the
two poles during this winding. For example, at the special time $t^*=3T/4$, the
wrapping arranges the FGS to pass over the north pole at some point $k_{\pi}$, and the gap in the ES is
closed.

In conclusion, we see that since the FGS ES is constructed from the time-dependent modes, the spectrum is
sensitive to the micro-motion of the states, and thus, unlike the quasi-energy spectrum, cannot rely on TRS arguments
that depend on integration over the full period. At most points during the drive TRS appears instantaneously
broken, except at $T/4$ and $3T/4$. The $\pi$ modes in general will ``flip" the spinor configuration, at $k = 0, \pi, k^*$,
which can be understood by the typical behavior of skyrmions under a band inversion. For
times away from the special TRS points, the FGS string on the Bloch-sphere is free to adjust to this flip by making a trivial loop on the sphere
leaving $|\mathbb{Z}_0 - \mathbb{Z}_\pi|$ windings. However at the two special TRS points (T/4,3T/4), the FGS string is forced to lie on
the $y-z$ great circle. Thus we have $|\mathbb{Z}_0 - \mathbb{Z}_\pi|$ for one of the TRS points and $|\mathbb{Z}_0 + \mathbb{Z}_\pi|$
for the other TRS point. Due to the nature of the drive chosen by us, the $|\mathbb{Z}_0 - \mathbb{Z}_\pi|$ winding is preserved throughout the period.
However other drives such as one that drives the pairing term, could break the invariant down to $\mathbb{Z}_2$ away from the TRS points.

\section{Entanglement of Physical or quenched state}\label{secQ}

We have mainly discussed the entanglement properties of the FGS.  We will now
discuss the entanglement of the physical state obtained from time-evolving under a
quench switch on protocol of the periodic drive. The entanglement properties of the
physical state are interesting to study as the ES will reveal topological features without
the need of introducing physical edges. Moreover unitary time evolution from some initial state
is much simpler than dissipative dynamics in the presence of a reservoir as the latter can create further complications in a driven system.

We will show that the quench ES will largely diverge from the FGS due to the presence of resonances.
Since the resonances involve a gap-closing process, the exact nature of the switch-on will not alter the
main conclusions of this section, as gap-closings do not have an adiabatic limit.

\begin{figure}
\includegraphics[width = .95\linewidth,keepaspectratio]{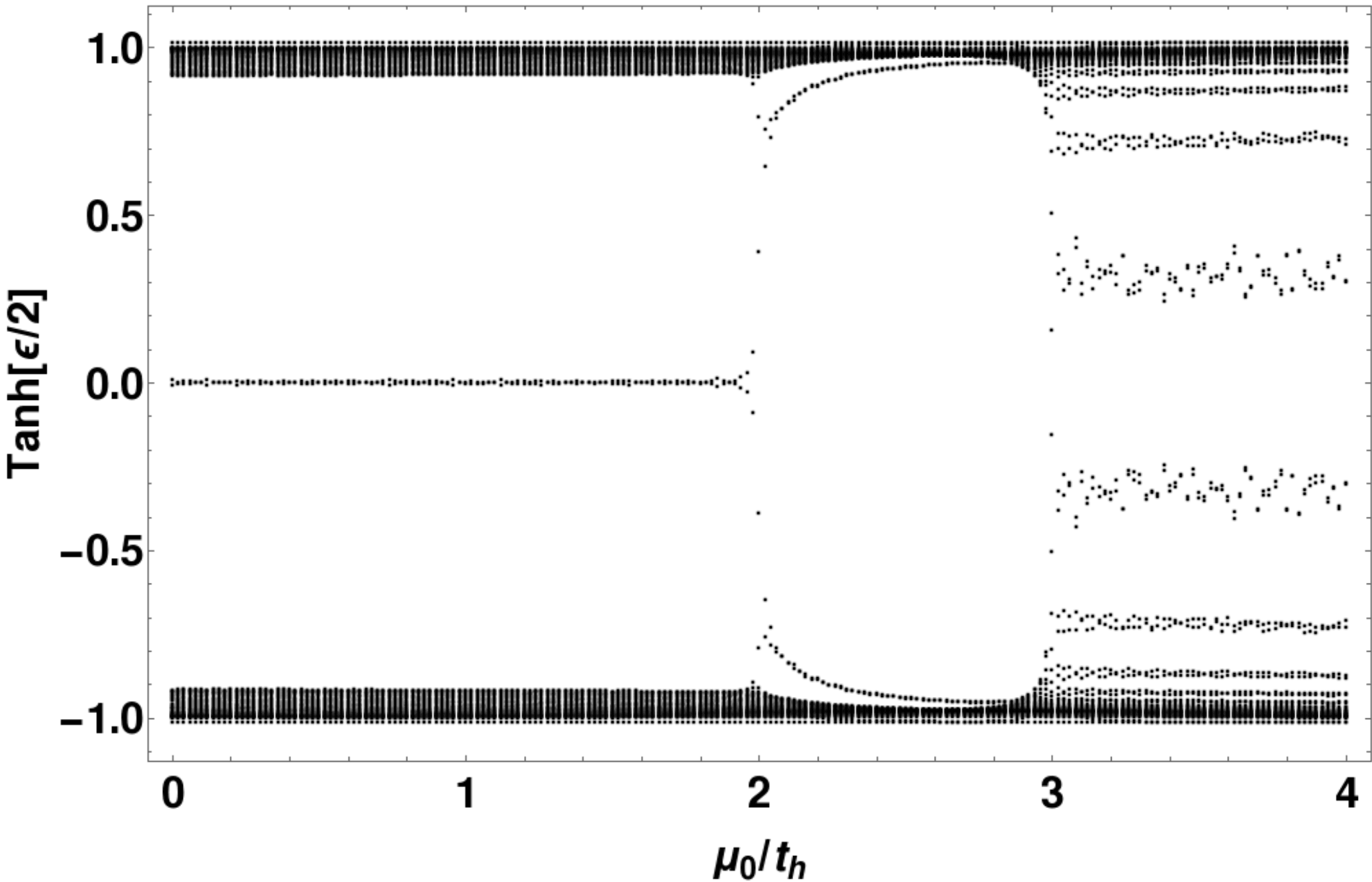}
\caption{ES for the quenched state at the start of a period
long after the quench, $\Omega/t_h = 5$, $\Delta/t_h = .5$, $\xi/t_h = 2.$, and $N = 35$.
This state is time evolved from the static ground state whose phase diagram corresponds to
a MZM phase ($0<\mu_0/t_h<2$) and a trivial phase ($2<\mu_0/t_h<4$). The state is evolved
according to the Floquet Hamiltonian, whose ground state has a MZM phase ($0<\mu_0/t_h<2$),
a trivial phase ($2<\mu_0/t_h<3$), and a MPM phase ($3<\mu_0/t_h<4$).
The MPM modes are absent in the ES of the physical state, and replaced by bulk excitations.
}
\label{om5q}
\end{figure}

\begin{figure}
\includegraphics[width = .95\linewidth,keepaspectratio]{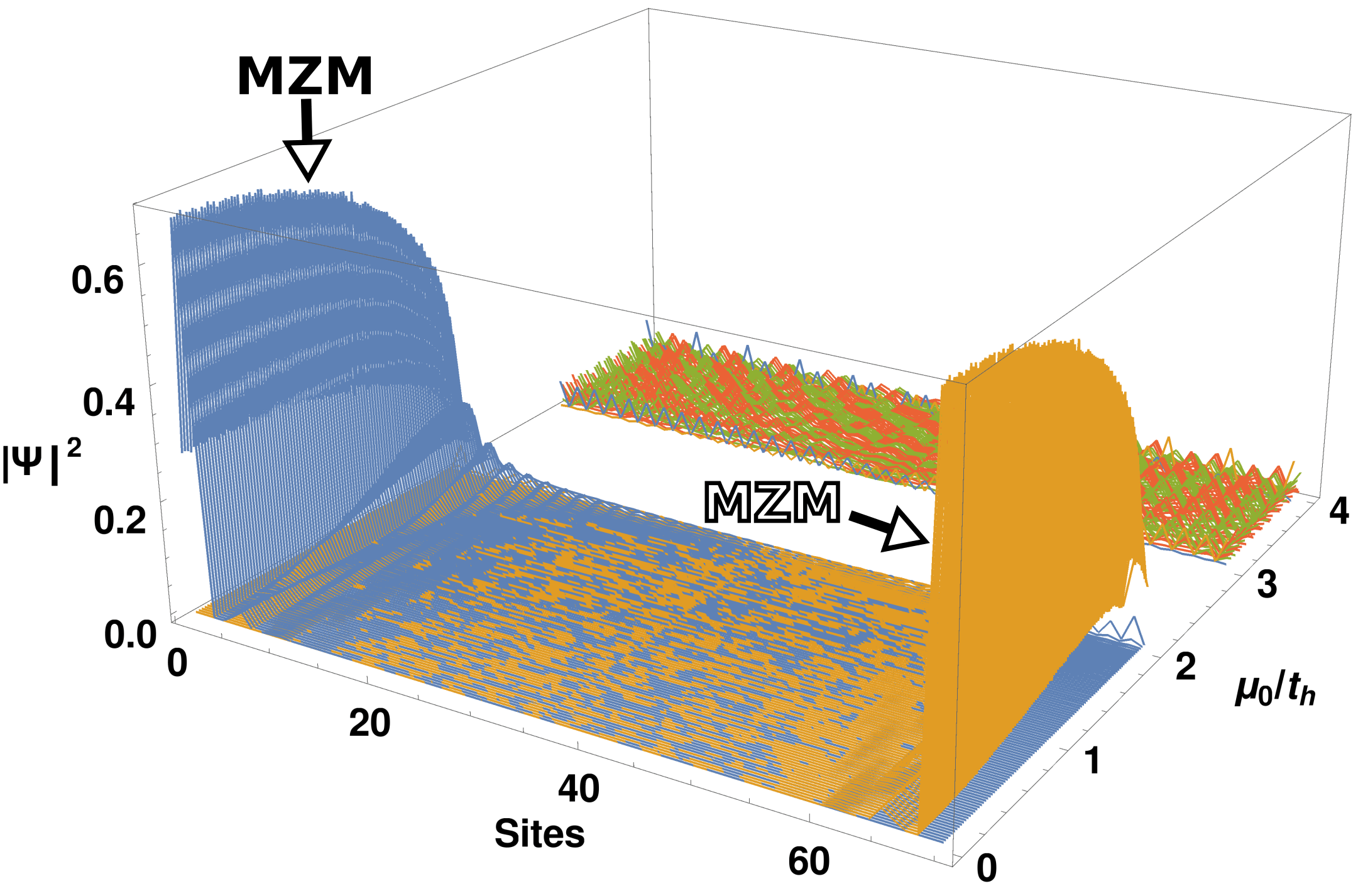}
\caption{ES edge Schmidt states for the quenched state at the start of a period
long after the quench, $\Omega/t_h = 5$, $\Delta/t_h = .5$, $\xi/t_h = 2$, and
$N = 35$.
Distinct colors (red, orange, green, blue) denote distinct eigenstates.
The MPM have merged with bulk excitations. The MZM on the other hand survive the quench.
}
\label{om5qe}
\end{figure}

\begin{figure}
\includegraphics[width = .95\linewidth,keepaspectratio]{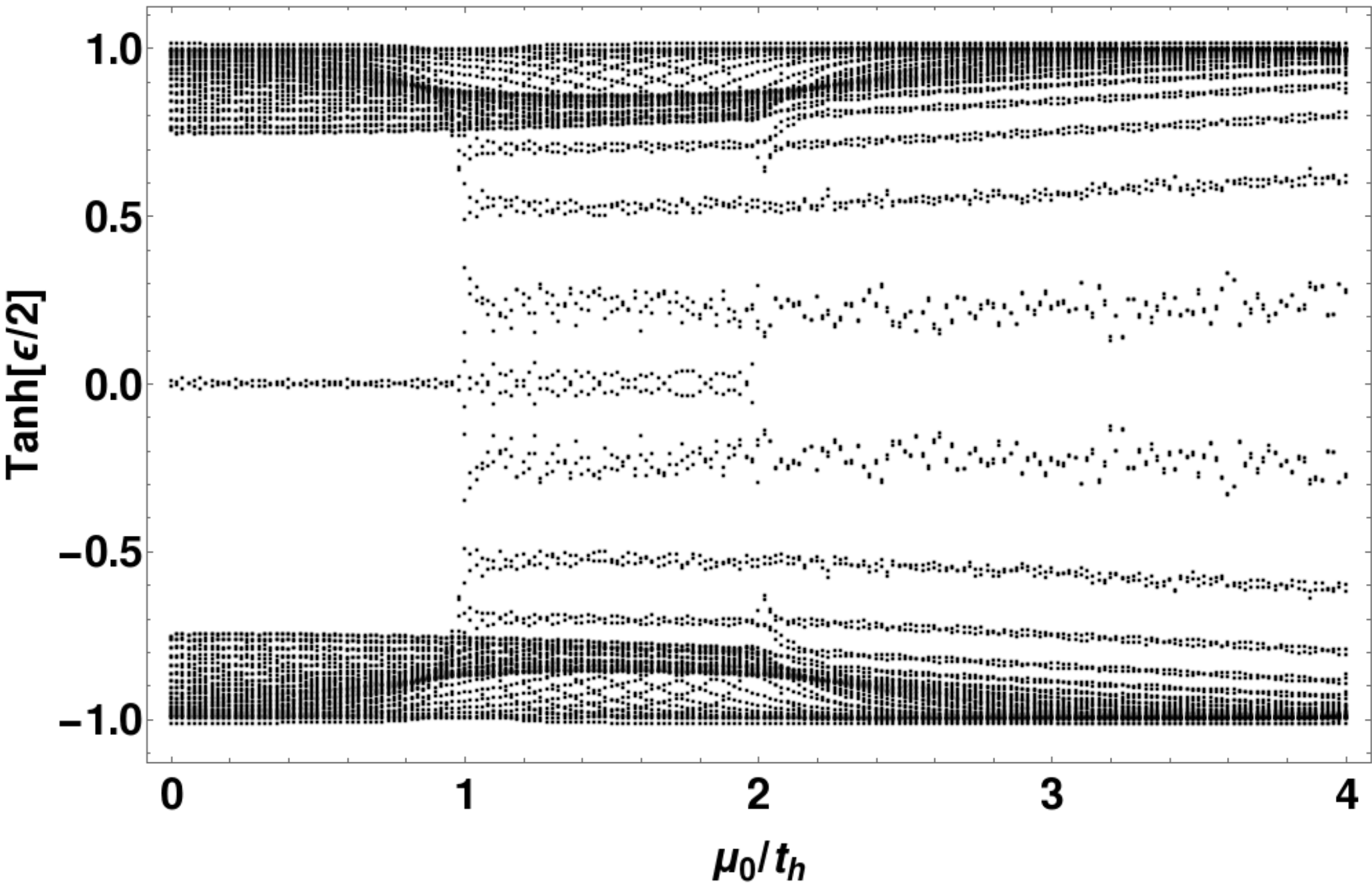}
\caption{
ES for the quenched state at the start of a period long after the quench,
$\Omega/t_h = 3$, $\Delta/t_h = .5$, $\xi/t_h = 2.$, and $N = 35$.
This state is time evolved from the static ground state whose phase diagram corresponds to
a MZM phase ($0<\mu_0/t_h<2$) and a trivial phase ($2<\mu_0/t_h<4$). The state is evolved
according to the Floquet Hamiltonian, whose ground state has a MZM phase ($0<\mu_0/t_h<1$),
a MZM\&MPM phase ($1<\mu_0/t_h<2$), and a MPM phase ($2<\mu_0/t_h<4$).
The MPM modes are absent in the ES of the physical state,
and the mid-gap states are bulk excitations. The MZM on the other hand are still visible.
}
\label{om3q}
\end{figure}

\begin{figure}[h!]
\includegraphics[width = .95\linewidth,keepaspectratio]{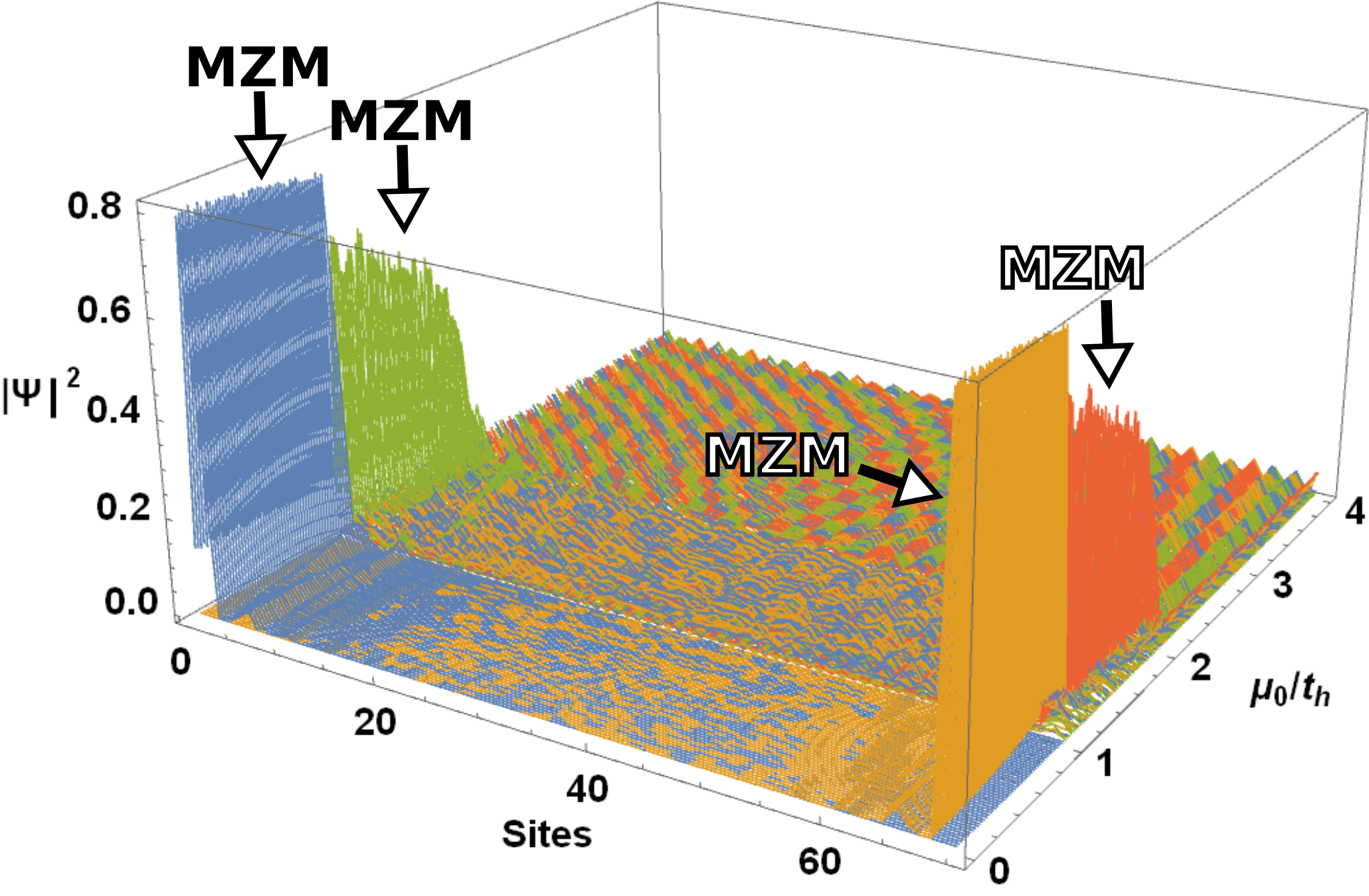}
\caption{ES edge Schmidt states for the quenched state at the start of a period
long after the quench, $\Omega/t_h = 3$, $\Delta/t_h = .5$, $\xi/t_h = 2$, and
$N = 35$. This figure only shows the states close to $\varepsilon = 0$ in the ES.
Distinct colors (red, orange, green, blue) denote distinct eigenstates.
Note the absence of the MPM modes.
}
\label{om3qe}
\end{figure}

For the quench, the ES is determined from equations \eqref{c_driven} and \eqref{f_driven} and inserted into
the Majorana correlation matrix, \eqref{corr}. The results are shown in figures \ref{om5q}, \ref{om3q} for the ES
and in figures \ref{om5qe} and \ref{om3qe} for the Schmidt states. Since we are only interested in the ES long times
after the quench occurred, terms with overlaps between the FGS and FES will yield zero after
taking the momentum integral. This is an example of dephasing that drives spatially extended quenched systems into
a diagonal ensemble. This also signifies that the ES at long times has no knowledge of
the quasi-energy spectrum and the time dependencies and topologies enter the ES
through the time-periodic Floquet modes.

First focusing on figures \ref{om5q} and \ref{om5qe}, we find that the off-resonant drive is largely equivalent
to what is in the FGS ES. There is an increase in the bulk excitations, but the entanglement gap
is still open. The resonant portion of the drive is however qualitatively new in the following manner. The physical state
fails to ``acquire" the MPM state and in its place we have highly excited bulk states, as expected \cite{Yates16}.
We expect for larger system sizes, the bulk states will completely fill the resonant portion of the ES, and  the
entanglement gap will close. The edge states in \ref{om5qe} shows the states in the resonant portion of the
ES are completely delocalized. In contrast, the off-resonant portion of the ES are only slightly modified from the FGS ES as we have
minimal bulk excitations in this portion of phase space.

Shifting attention to figures \ref{om3q} and \ref{om3qe}, the off-resonant region is largely the same
as the FGS. The region where we have a MPM phase in the FGS ES, corresponds to highly excited bulk states and no MPM,
just as in the $\Omega/t_h = 5$ case discussed above. The new region is the MZM\&MPM phase in the FGS. In the physical state
the ES shows a central edge state surviving the quench, and the appearance of highly excited bulk states. Essentially,
the physical state fails to acquire the $\pi$ modes, so there is no issue of the MZM\&MPM gapping
itself out and ES maintains the original (pre-quench) MZM topology of the state.
Looking at figure \ref{om3qe}, the
Schmidt states for those levels are indeed edge states and remain sharply defined throughout the drive.
Even though the central MZM persists in the physical state, it is not topologically robust because
the gap between the topological and bulk states in the MZM\&MPM phase can become smaller in the limit of larger entanglement cuts.

To summarize, we find that the topology of the physical state is that of the initial state before unitary time-evolution.
However, these inherited edge modes have weaker topological
protection due to nearby bulk excitations created when the drive is resonant.

\section{Discussion}\label{secD}
We investigated the entanglement properties of a periodically driven Kitaev chain through the use of the
entanglement spectrum (ES). The goal was to better understand the topological features of the eigenstates of a driven system in a general
way without resorting to making assumptions of thermal equilibrium occupation of the states. To this end we studied the entanglement properties
of the exact eigenstate of the Floquet Hamiltonian,
which we call the Floquet Ground State (FGS), and also that of a physical state
arising from unitary time-evolution following a quench of the periodic drive.

We made use of a Majorana correlation matrix in order to construct the reduced density matrix, and the corresponding entanglement Hamiltonian.
We carried this out for three parent states, one was the ground state of the static Hamiltonian, the second was the FGS, and the third was the
physical state. Both
numerical and analytical arguments were presented, demonstrating the bulk-boundary correspondence that exists for the
entanglement Hamiltonian, where the boundary is now that of a fictitious entanglement cut.

While the ES provides topological information about the state in question, we found that the
$\mathbb{Z}_0 \times \mathbb{Z}_\pi$ topological index for the Floquet Hamiltonian does not carry over to the entanglement
Hamiltonian. In particular the number of topological Majorana modes in the ES vary during the period of the drive in a way such that 
at the two time reversal symmetric points of the drive ($t^*=T/4, 3T/4$ in our example), the number of Majorana modes in the ES are
$|\mathbb{Z}_0 - \mathbb{Z}_\pi|$ and $|\mathbb{Z}_0 + \mathbb{Z}_\pi|$ respectively. 
The essential reasoning behind this breakdown is that the ES is not a periodic
quantity and thus the topological states all reside at zero entanglement energy, while the $\mathbb{Z}\times\mathbb{Z}$ classification
required an energy separation for the $0$ and the $\pi$ modes. Thus now one may couple some of the MZM and
MPM modes in the ES, reducing the number of Majorana modes.

We also considered NNN hopping in order to generate more edges states to further
demonstrate this result.
The topology of the Floquet ground state and the corresponding ES was made explicit through usage of Bloch-sphere diagrams showing that
the FGS spinor has well defined winding only at $t^*$ where the $|\mathbb{Z}_0 \pm \mathbb{Z}_\pi|$ Majorana modes appear. The persistence of
$|\mathbb{Z}_0 - \mathbb{Z}_\pi|$ away from $t^*$ despite the lack of a well defined winding of the spinor, was because the periodic
drive was applied to the chemical potential. Periodic driving to the pairing on the other hand will reduce the Majorana modes in the ES at times
other than $t^*$ to $\mathbb{Z}_2$.

When the MZM\&MPM modes couple, one obtains complex or Dirac fermions with entanglement energies lifted away from zero, 
but with Schmidt states still localized at the entanglement cut. However unlike accidental edge Dirac
fermions that can always occur even in the trivial phase, the ones that arise in the topologcal phase are protected because 
the Dirac fermions again have to uncouple into two Majorana fermions at the
special TRS points $t^*$ of the drive.

The study of the ES of the physical state, and the corresponding Schmidt states show that
the $\pi$ modes do not appear in the ES. This is because they co-occur with large number of bulk excitations and thus hybridize with them.
In contrast the MZMs are still visible, where they are inherited from the wavefunction before the quench~\cite{Rigol14b}.
However we expect that these modes lose their topological protection due to nearby bulk excitations created by the resonant laser~\cite{Yates16}.

In a study of the entanglement properties of the Floquet Chern Insulator, it was found that~\cite{Yates16}
for a physical state obtained from taking half-filled graphene, and time-evolving it with a circularly polarized
periodic drive, chiral modes appeared in the ES of the physical state even though they are absent in the initial state.
This is in contrast to what we find here, where only the topological
modes of the initial state persist during the time-evolution. This difference is due to the fact that graphene, being a semi-metal,
is in a topologically critical state, and has its
own edge-states. These edge-states easily acquire a chiral behavior under a laser quench.

{\sl Acknowledgements:}
The authors thank Yonah Lemonik for many useful discussions.
This work was supported by the US Department of Energy,
Office of Science, Basic Energy Sciences, under Award No.~DE-SC0010821.

%

\end{document}